\documentclass[twocolumn]{aastex631}

\usepackage{multirow}
\usepackage[flushleft]{threeparttable}
\usepackage{subfigure}

\newcommand{\DSTA}{DS\,Tuc\,A}
\newcommand{\DSTB}{DS\,Tuc\,B}

\begin{document}

\title{Stellar X-ray variability and planetary evolution in the DS\,Tucanae system}

\author[0000-0002-3641-6636]{George W. King}
\affiliation{Department of Astronomy, University of Michigan, Ann Arbor, MI 48109, USA}
\affiliation{Department of Physics, University of Warwick, Gibbet Hill Road, Coventry, CV4 7AL, UK}
\affiliation{Centre for Exoplanets and Habitability, University of Warwick, Gibbet Hill Road, Coventry, CV4 7AL, UK}

\author[0000-0002-5466-3817]{L\'{i}a R. Corrales}
\affiliation{Department of Astronomy, University of Michigan, Ann Arbor, MI 48109, USA}

\author[0000-0002-9148-034X]{Vincent Bourrier}
\affiliation{Geneva Observatory, University of Geneva, Chemin Pegasi 51b, CH-1290 Versoix,  Switzerland }

\author[0000-0002-2248-3838]{Leonardo A. Dos Santos}
\affiliation{Space Telescope Science Institute, 3700 San Martin Drive, Baltimore, MD 21218, USA}

\author[0000-0002-9365-2555]{Lauren Doyle}
\affiliation{Department of Physics, University of Warwick, Gibbet Hill Road, Coventry, CV4 7AL, UK}
\affiliation{Centre for Exoplanets and Habitability, University of Warwick, Gibbet Hill Road, Coventry, CV4 7AL, UK}

\author[0000-0001-8884-9276]{Baptiste Lavie}
\affiliation{Geneva Observatory, University of Geneva, Chemin Pegasi 51b, CH-1290 Versoix,  Switzerland }

\author[0000-0001-8722-9710]{Gavin Ramsay}
\affiliation{Armagh Observatory and Planetarium, College Hill, Armagh, N Ireland, BT61 9DG, UK}

\author[0000-0003-1452-2240]{Peter J. Wheatley}
\affiliation{Department of Physics, University of Warwick, Gibbet Hill Road, Coventry, CV4 7AL, UK}
\affiliation{Centre for Exoplanets and Habitability, University of Warwick, Gibbet Hill Road, Coventry, CV4 7AL, UK}



\begin{abstract}

We present an analysis of four \textit{Chandra} observations of the 45\,Myr old DS\,Tuc binary system. We observed X-ray variability of both stars on timescales from hours to months, including two strong X-ray flares from star A. The implied flaring rates are in agreement with past observations made with \textit{XMM-Newton}, though these rates remain imprecise due to the relatively short total observation time. We find a clear, monotonic decline in the quiescent level of the star by a factor 1.8 across eight months, suggesting stellar variability that might be due to an activity cycle. If proven through future observations, \DSTA\ would be the youngest star for which a coronal activity cycle has been confirmed. The variation in our flux measurements across the four visits is also consistent with the scatter in empirical stellar X-ray relationships with Rossby number. In simulations of the possible evolution of the currently super-Neptune-sized planet \DSTA\,b, we find a range of scenarios for the planet once it reaches a typical field age of 5\,Gyr, from Neptune-size down to a completely stripped super-Earth. Improved constraints on the planet's mass in the future would significantly narrow these possibilities. We advocate for further \textit{Chandra} observations to better constrain the variability of this important system.

\end{abstract}




\section{Introduction}
\label{sec:intro}

In recent years, an increasing number of planets have been discovered transiting young stars, driven by surveys of open clusters and associations by the \textit{K2} \citep[e.g.][]{David2016,Libralato2016,Mann2017,Mann2018} and \textit{TESS} missions \citep[e.g.][]{Bouma2020,Rizzuto2020,Wood2023}. Characterizing this growing population of young planets offers a tantalizing pathway to understanding the evolution of planets orbiting close-in to their star. For planets smaller than Jupiter-size, there is already evidence that young planets are larger than their older counterparts \citep[e.g.][]{Mann2017,Tofflemire2021}.

The shrinkage of planetary radii between early and field ages is expected. First, planets cool following formation, leading to a steady shrinkage of their atmospheres. Second, X-ray and extreme-ultraviolet (EUV, together XUV) irradiation is thought to drive substantial atmospheric escape from the atmospheres of close-in exoplanets \citep[e.g.][]{Lammer2003,M-C2009,Owen2012}. For Neptune-sized planets and smaller, the escape can be severe enough to significantly evolve the mass and radius of the planet, with removal of the entire primordial envelope possible in the most extreme cases \citep[e.g.][]{Lopez2012,Owen2013}. 
The radiation of energy from the planet's core can also drive atmospheric escape for sub-Neptune-sized planets, potentially further reducing the mass and radius of the planet \citep[e.g.][]{Ginzburg2018,Gupta2019}. The timescale on which cores can be stripped by this ``core-powered mass loss" is 1-2\,Gyr \citep{Gupta2020}. This is much longer than the typical timescale for atmosphere stripping by XUV photons, which has been widely suggested to be a few hundred Myr \citep[e.g.][]{Lopez2013,Lammer2014,Owen2017}.

The timescale on which XUV-driven escape can significantly alter the planetary atmosphere is largely driven by the time evolution of stellar XUV emission. For FGK stars, the ratio of the X-ray and bolometric luminosities, $L_{\rm x}/L_{\rm bol}$, remains at a characteristic level of $\sim 10^{-3}$ across the first 100\,Myr or so, due to the saturation of magnetic dynamo. After this period, $L_{\rm X}/L_{\rm bol}$ falls off with a power law behaviour as magnetic braking slows the rotation of the star, thus weakening the dynamo over time \citep{Skumanich1972,Kawaler1988}. There is evidence that the decline in the ratio for EUV emission, $L_{\rm EUV}/L_{\rm bol}$, drops off at a slower rate than for X-ray \citep[e.g.][]{EUVevolution,Johnstone2021}, possibly lengthening the XUV evolutionary timescale. The spindown and thus XUV decline in M dwarfs is also thought to be substantially slower \citep[e.g.][]{Newton2016,Pass2022,R-Y2023,Engle2024}. However, in general, it is still the very youngest exoplanetary systems that offer much promise for the observation and characterization of planetary escape and evolution.

Outstanding questions also still remain over the variation of XUV emission, which is important to characterize as this could result in changes to the escape rate \citep[e.g.][]{LDE2012,Bourrier2020}. Empirical relationships for $L_{\rm X}/L_{\rm bol}$ with both age and rotation typically show a scatter of an order of magnitude or more around the best fit \citep[e.g.][]{Pizzolato2003,Jackson2012,Wright2011}. This is true even for the early-time saturated regime, where the emission is often assumed to be a single $10^{-3}$ value. There are two possibilities for the generation of this scatter. First, there could be intrinsic scatter in the saturation level from star-to-star, and/or the time evolution of the X-ray emission thereafter. In the second mechanism, the time-averaged emission is the same, but there exists variation about the mean $L_{\rm X}/L_{\rm bol}$ over time. Such variation could result from several phenomena, for example stochastic changes as active regions appear and disappear. Activity cycles akin to the 11-year Solar cycle also result in variation about the mean value, with the scatter then resulting from measurements being very short snapshots within that cycle. Flares are typically easier to account for, but this is not always (or only imperfectly) done. 
Investigation is necessary to determine which, if either, dominates the generation of the scatter.

Very few stars have had a coronal activity cycle successfully measured, with the youngest being the 440\,Myr K dwarf $\epsilon$\,Eri, which shows a 2.9\,yr cycle \citep{Coffaro2020}. Attempts have been made to detect cycles for younger stars \citep[e.g.][]{Lalitha2013,Coffaro2022,Maggio2023}, but none have so far shown evidence of cyclical behavior. The few detections for older stars show a possible relationship where stars with a lower Rossby number (i.e. a faster rotation rate, and therefore younger stars) appear to exhibit smaller cycle amplitudes. This effect is thought to be caused by a high magnetic filling factor, which leaves little room for further activity enhancement \citep{Wargelin2017,Coffaro2022}.

In this work, we examine the X-ray activity of DS\,Tuc, a binary star system in the 45\,Myr Tucana-Horologium young moving group. The two stellar components are a G6V primary and K3V secondary \citep{Torres2006}, separated by 5$"$ on the sky (220\,AU). The discovery by \textit{TESS} of a 5.70\,R$_\oplus$ planet, between the sizes of Neptune and Saturn, orbiting the primary star \DSTA\ was reported by \citet{Newton2019} and \citet{Benatti2019}. We adopt the \citet{Newton2019} parameters for the system in this work, unless otherwise stated. There are no planets of this intermediate size in the Solar System, and relatively few among the exoplanet population too. Given the very young age of the system, significant XUV irradiation of \DSTA\,b is likely ongoing. This could mean that by the time the system reaches field ages of gigayears, the planet may lose sufficient mass that its radius shrinks down to join the large population of (sub-)Neptune-sized planets uncovered by \textit{Kepler} and \textit{TESS}.

\begin{figure*}
\centering
 \includegraphics[width=\textwidth]{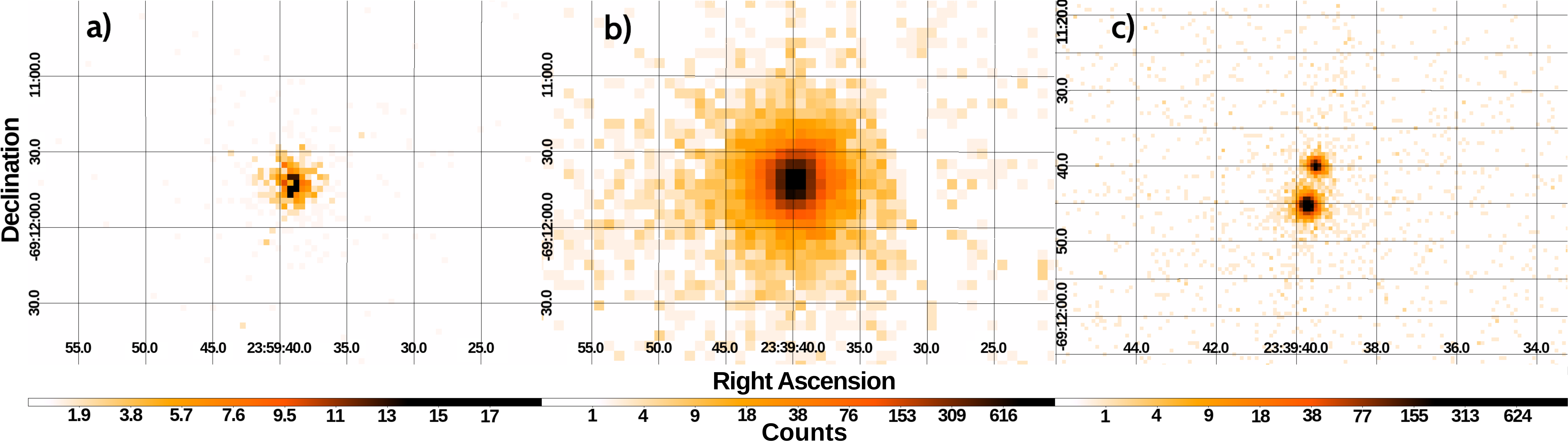}
 \caption{X-ray images from a) \textit{Swift} XRT, b) \textit{XMM-Newton} EPIC-pn, and c) \textit{Chandra} ACIS-S, highlighting how our new \textit{Chandra} data fully separates \DSTA\ (bottom star) and \DSTB\ (top star) spatially in X-rays for the first time. The \textit{XMM-Newton} image is from obsID 0863400901 (previously analyzed by \citealt{Benatti2021}). The \textit{Chandra} image is from obsID 25103. Note the different axes on the \textit{Chandra} image, as it is zoomed in compared to the other two. The colorbar scales are also different between the images, as the number of counts per pixel in the center of the PSF varies greatly between the observations.}
 \label{fig:imageComp}
\end{figure*}

In order to characterize the high-energy environment experienced by \DSTA\,b, two previous observations of the system have been made with \textit{XMM-Newton} (ObsIDs: 0863400901, PI S. Wolk; 0864340101, PI A. Maggio). The small separation of the two stars on the sky leads to a significant overlap in their point spread functions (PSFs; see Fig. \ref{fig:imageComp}) in the European Photon Imaging Cameras \citep[EPIC;][]{Struder2001,Turner2001} of \textit{XMM-Newton}. Nonetheless, \citet{Benatti2021} were able to use the first of these observations to estimate the respective contributions of the two stars to the overall emission and obtain coronal properties for both. In their analysis of the second observation, \citet{Pillitteri2022} reported two X-ray flares originating from \DSTA, noting a delay in the flare peak in X-rays versus near-UV, and discuss the possible effects of the flares on the planet. They additionally assessed data from the \textit{XMM-Newton} Reflection Grating Spectrometer (RGS) instruments, identifying and quantifying line strengths of the emission in the quiescent and flaring states.

In this work, we present four new \textit{Chandra} observations of the DS\,Tuc system in order to characterize the stellar XUV emission, its variability, and its effect on the planet now and in the future. The major advantage of using \textit{Chandra} to measure the X-ray emission of DS\,Tuc is the superior, 0.5$"$ spatial resolution afforded by the telescope. This means the two stars are clearly separated in the images, allowing for a much more accurate assessment of their relative emissions in the observations. We also analyze a smaller dataset obtained with the Neil Gehrels \textit{Swift} Observatory, though the PSF of its X-ray Telescope (XRT) is substantially lower than that of \textit{Chandra}.

We look at the \textit{Chandra} data in Section~\ref{sec:Chandra}, and the \textit{Swift} data in Section~\ref{sec:Swift}. We take a closer look at flares, longer term variation, and compare our results with empirical X-ray relations in Sections~\ref{sec:flares}, \ref{sec:temporal}, and \ref{sec:compRel}, respectively. The quiescent luminosity is then used as an anchor point in simulating the possible lifetime evolution and current mass loss of \DSTA\,b in Section~\ref{sec:planetEff}, before some concluding remarks in Section~\ref{sec:conclusions}

\section{\textit{Chandra} observations and results}
\label{sec:Chandra}

\begin{table}
\centering
\caption{Observation log for our four \textit{Chandra} observations.}
\label{tab:obs}
\begin{tabular}{cccc}
\hline
Obs ID & Start Time       & Livetime & Planet \\
       & (TDB)            & (ks)     & phase  \\
            \hline
25103  & 2022-02-20T22:27 & 13.64    & 0.6332 -- 0.6552       \\
25104  & 2022-05-01T20:18 & 13.95    & 0.2228 -- 0.2448       \\
25105  & 2022-07-11T04:16 & 15.34    & 0.8656 -- 0.8898       \\
25106  & 2022-10-12T15:46 & 13.95    & 0.3513 -- 0.3733       \\
\hline
\end{tabular}
\tablecomments{A planet phase of 0 corresponds to the center of primary transit of \DSTA\,b. TDB is the Barycentric Dynamical Time system.}
\end{table}

We observed the DS\,Tuc system with \textit{Chandra} Advanced CCD Imaging Spectrometer S array (ACIS-S) four times between February and October 2022. Our observation log is given in Table \ref{tab:obs}. In each observation, we used only the S3 chip, and operated it in 1/8 subarray mode. We chose this mode in order to set the frame time to the minimum 0.4\,s and avoid pile-up issues.

In Figure \ref{fig:imageComp}, we display a comparison of X-ray images of the DS\,Tuc system from three instruments: \textit{Swift} XRT, \textit{XMM-Newton} EPIC-pn, and \textit{Chandra} ACIS-S. This figure clearly demonstrates the advantage of exploiting the superior spatial resolution of \textit{Chandra} for observing DS\,Tuc in X-rays, with the observations we present in this paper fully separating the PSF of the two stars for the first time at these wavelengths. We note that the EPIC-pn PSF appears larger than the XRT PSF. This is because of the vastly smaller number of detected counts for XRT making it appear smaller compared to the background.

In our analyses, we used \texttt{Ciao} 4.14 \citep{Fruscione2006}, and in our data reductions we followed the \texttt{Ciao} science threads\footnote{\url{https://cxc.cfa.harvard.edu/ciao/threads/}}. We used a 3$"$ radius source extraction regions for \DSTA\ and 2.5$"$ regions for \DSTB\ in extracting the light curves and spectra that we describe in the following subsections. For the background assessment, we used two 30$"$ regions approximately 1$'$ away on either side of the binary.

\subsection{X-ray light curves}
\label{ssec:xLC}

\begin{figure*}
\centering
 \subfigure[0.5--10\,keV light curve]{\includegraphics[width=0.41\textwidth]{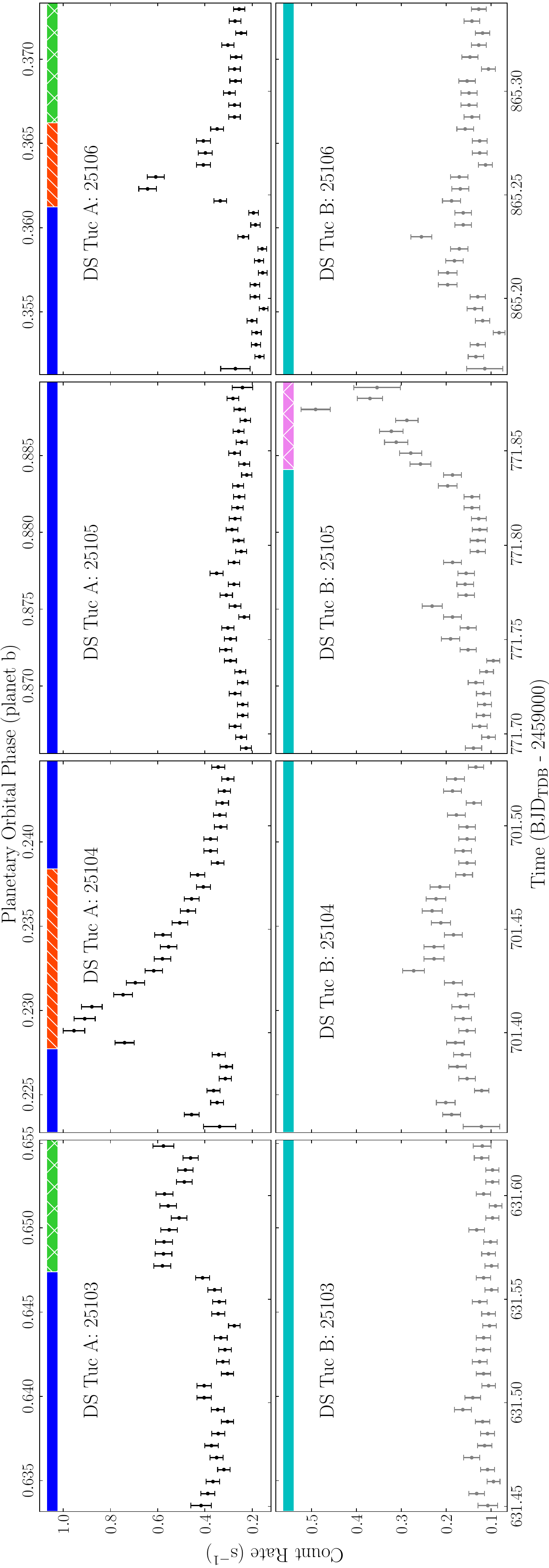}
 \label{fig:xLC}}
 \subfigure[Hardness ratio plot]{\includegraphics[width=0.41\textwidth]{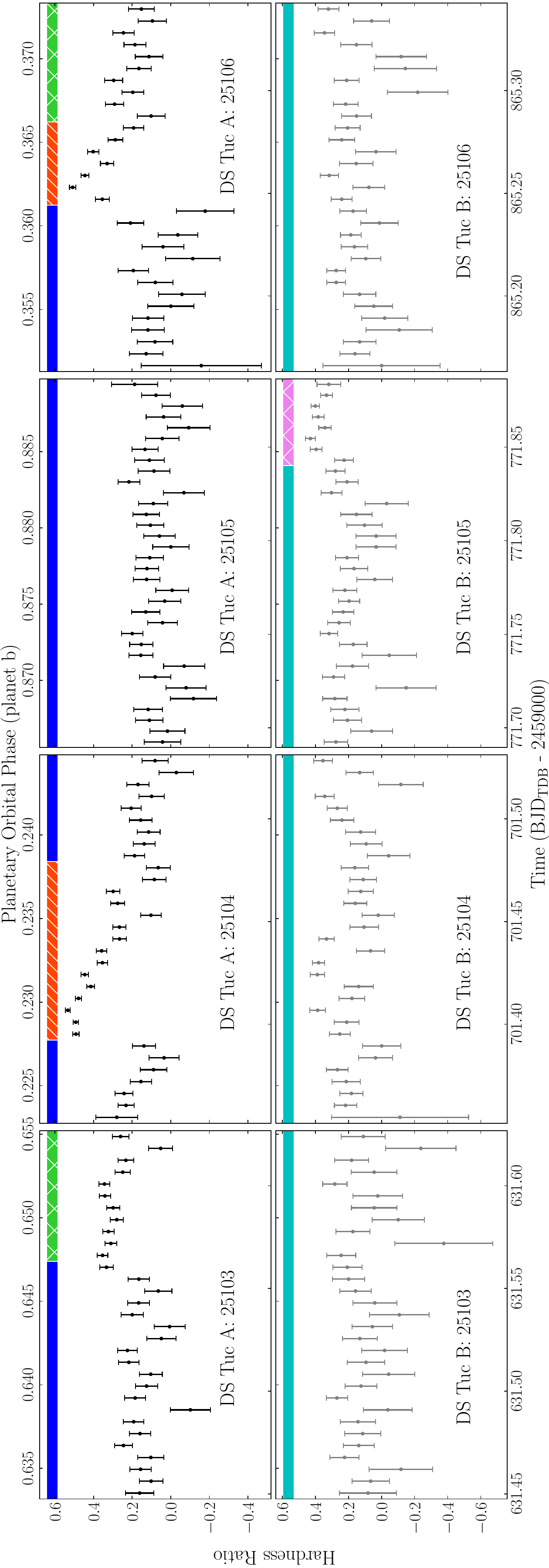}
 \label{fig:xHR}}
 \caption{Chandra ACIS-S X-ray light curves and hardness ratios (see definition in the main text). In both subfigures, the points for \DSTA\ are shown in the top panels with the black points, and for \DSTB\ in the bottom panels with gray points. Epochs of the two flares on star A are displayed with the red single-hatched ribbon at the top of the relevant panel. Other epochs defined to have elevated count rate are shown with a cross-hatched ribbon (green for star A, magenta for star B). All other epochs were defined as quiescent, and are shown with a solid ribbon (blue for star A, cyan for star B). These definitions are used to define separate spectra extractions in Section \ref{ssec:xSpec}, and the ribbon colors are the same as those used for the spectral points in Fig~\ref{fig:xSpec}.}
 \label{fig:xLC+HR}
\end{figure*}



In Figure \ref{fig:xLC} we display the 0.5 -- 10\,keV light curve in each of the four \textit{Chandra} observations for both stars, binned to 500\,s cadence. As with the second \textit{XMM-Newton} observation analyzed by \citet{Pillitteri2022}, we detected two clear flares from \DSTA\ in our observations: one particularly strong flare in observation 25104, and another in 25106. Star A shows a further count rate enhancement at the end of observation 25103, however the light curve profile is not consistent with a typical flare. DS\,Tuc\,B also shows an intriguing, upward ramp in 25105, but the end of the observation means we have no information about whether there was a flare-like decay. The rise up in count rate also appears slower than the two bona fide flares. There are also other variations in the count rate of star B (notably in observations 25104 and 25106) that could be due to smaller flaring events, but within the noise level of the data we cannot be conclusive.

Past observations of coronal X-ray flares have shown a characteristic hardening of the emission during the flare \citep[e.g.][]{Reale2001,Telleschi2005,Pye2015}. We examine this for our flares by calculating hardness ratios, $HR$, for each time bin in the light curves which we plot in Figure \ref{fig:xHR}. We define the hardness ratio as 
\begin{equation}
    HR = \frac{H-S}{H+S}
    \label{eq:HR}
\end{equation}
where $S$ and $H$ are the soft band (0.5--1.25\,keV) and hard band (1.25--10\,keV) count rates. The hardness ratio plot reveals a significant hardening of the emission in both flares of star A, in line with expectations. The two other possible flaring events for star A in 25103 and star B in 25105 also appear to show a hardening of the emission at the time that the count rate becomes substantially elevated above quiescence. This does add considerable weight towards these events being flares, however the lack of clear decay for the star A event, and the lack of any post-peak data for the star B event are such that we designate these epochs as ``elevated" through the rest of the paper. 
We also designate the post flare data for star A in 25106 as ``elevated", since the constant level that the emission returns to is higher than the pre-flare quiescent count rate.

In Section \ref{ssec:xSpec}, we extract separate spectra for flares and elevated epochs. The end of each flare was determined by eye as the point where the count rate returns to its pre-flare level. Our definitions are displayed on Figure \ref{fig:xLC}, with different color ribbons denoting the different stellar state definitions. 
We also discuss the flares and the relative contributions of each type of epoch for star A further in Section \ref{sec:flares}.


\subsection{X-ray spectra}
\label{ssec:xSpec}

\begin{figure*}
\centering
 \includegraphics[width=\textwidth]{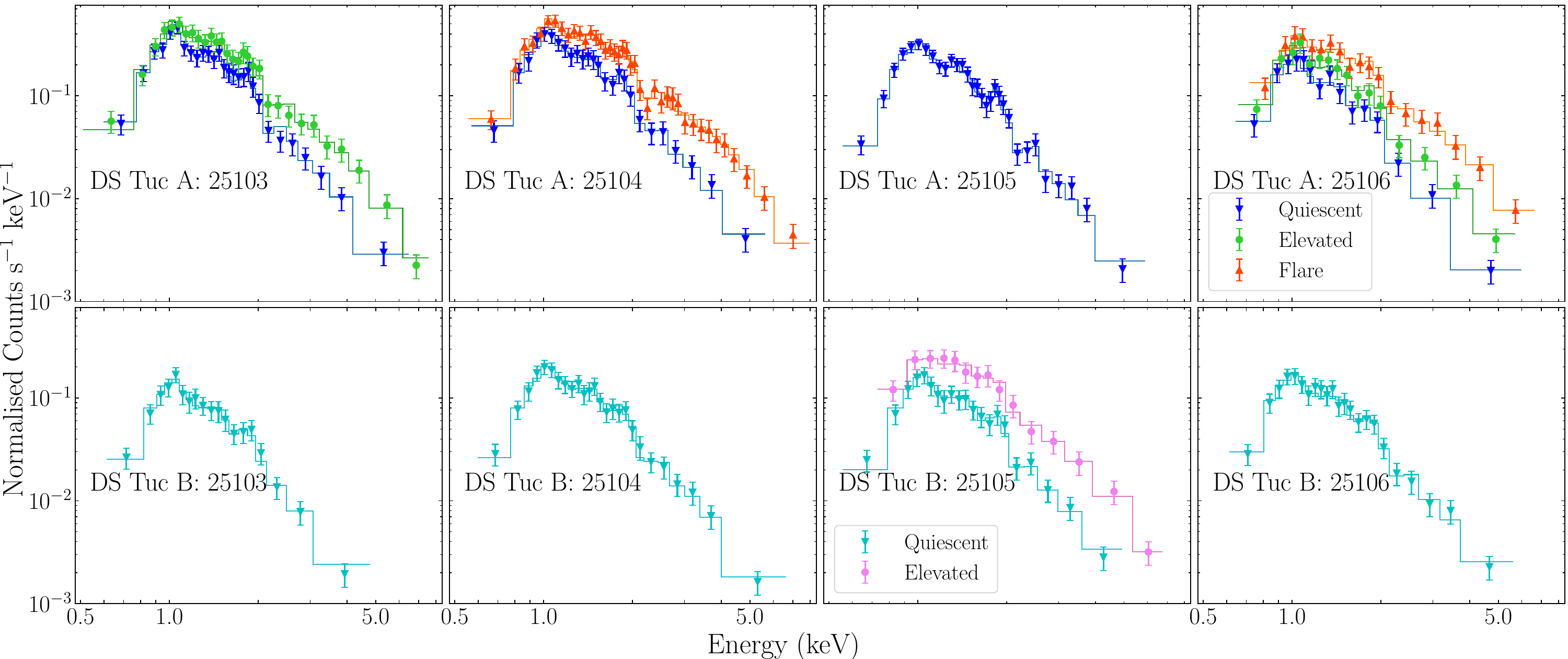}
 \caption{ACIS-S spectra for \DSTA\ (top panels) and B (bottom panels) in each of the four observations. The spectra have been extracted and fitted separately for quiescent, elevated, and flaring periods, according to the definitions displayed in Fig. \ref{fig:xLC}. The overplotted histograms show the best fit model to each spectrum.}
 \label{fig:xSpec}
\end{figure*}

\begin{table*}
\centering
\caption{Best fit plasma temperatures and abundances for \DSTA\ and B from our spectral fits.}
\label{tab:temp+abund}
\begin{tabular}{ccccccccc}
\hline
Star     & $kT_1$                       & $kT_2$                    & $kT_3$                    & \multicolumn{5}{c}{Abundances$^\dagger$}                                                                                                                                                                              \\
         & (keV)                     & (keV)                  & (keV)                  & O                                       & Ne                                      & Mg                                      & Si                                         & Fe                                         \\ \hline
\DSTA\ & $0.757^{+0.024}_{-0.022}$ & $1.56^{+0.18}_{-0.19}$ & 7.71$^\ddagger$        & \multirow{2}{*}{$3.67^{+0.92}_{-0.77}$} & \multirow{2}{*}{$3.14^{+0.58}_{-0.50}$} & \multirow{2}{*}{$0.75^{+0.12}_{-0.11}$} & \multirow{2}{*}{$0.505^{+0.072}_{-0.064}$} & \multirow{2}{*}{$0.587^{+0.084}_{-0.074}$} \\
\DSTB & $0.770^{+0.029}_{-0.031}$ & $1.49^{+0.13}_{-0.10}$ & $3.13^{+0.44}_{-0.32}$ &                                         &                                         &                                         &                                            &                                            \\ \hline
\end{tabular}
\tablecomments{$^\dagger$ Abundances are wrt Solar \citep{Asplund2009}. $^\ddagger$ The highest temperature for \DSTA\ was unconstrained at the upper end, running up to the hard limit - the lower 1-$\sigma$ confidence interval was 4.75 keV.}
\end{table*}

\begin{table*}
\centering
\caption{Best fit emission measures, fluxes and luminosities for each extracted spectrum.}
\label{tab:EM+fluxes}

\begin{tabular}{ccccccccccccc}
\hline
\# & Star & Obs   & Stellar & EM$_{1}$             & EM$_{2}$             & EM$_{3}$             & $F_{\rm x, 0.5}$           & $L_{\rm x, 0.5}$        & $\frac{L_{\rm x, 0.5}}{L_{\rm bol}}$ & $F_{\rm x, 0.1}$       & $L_{\rm x, 0.1}$        & $\frac{L_{\rm x, 0.1}}{L_{\rm bol}}$ \\
   &      &       & State   & ($a$)                & ($a$)                & ($a$)                & ($b$)                      & ($c$)                   & ($\times 10^{-4}$)                   & ($b$)                  & ($c$)                   & ($\times 10^{-4}$)                   \\ \hline
1  & A    & 25103 & Q       & $34.3^{+4.9}_{-4.4}$ & $36.9^{+4.5}_{-6.0}$ & $6.3^{+5.5}_{-4.0}$  & $5.618^{+0.094}_{-0.218}$  & $13.12^{+0.22}_{-0.51}$ & $4.73^{+0.12}_{-0.20}$               & $5.89^{+0.38}_{-0.45}$ & $13.75^{+0.90}_{-1.04}$ & $4.95^{+0.33}_{-0.39}$               \\
2  & A    & 25103 & E       & $33.8^{+5.7}_{-4.9}$ & $49^{+13}_{-15}$     & $36^{+15}_{-10}$     & $8.68^{+0.13}_{-0.42}$     & $20.26^{+0.31}_{-0.99}$ & $7.30^{+0.17}_{-0.38}$               & $7.72^{+0.54}_{-0.58}$ & $18.0^{+1.3}_{-1.4}$    & $6.50^{+0.47}_{-0.50}$               \\
3  & A    & 25104 & Q       & $32.1^{+4.9}_{-4.3}$ & $38.0^{+5.1}_{-6.6}$ & $7.0^{+6.0}_{-4.8}$  & $5.55^{+0.11}_{-0.23}$     & $12.96^{+0.25}_{-0.54}$ & $4.67^{+0.12}_{-0.21}$               & $5.76^{+0.39}_{-0.45}$ & $13.45^{+0.90}_{-1.06}$ & $4.85^{+0.34}_{-0.39}$               \\
4  & A    & 25104 & F       & $34.9^{+5.5}_{-4.8}$ & $45^{+17}_{-21}$     & $60^{+20}_{-13}$     & $10.478^{+0.099}_{-0.518}$ & $24.47^{+0.23}_{-1.21}$ & $8.82^{+0.18}_{-0.46}$               & $8.56^{+0.57}_{-0.59}$ & $20.0^{+1.3}_{-1.4}$    & $7.20^{+0.49}_{-0.52}$               \\
5  & A    & 25105 & Q       & $37.9^{+5.0}_{-4.5}$ & $15.4^{+3.2}_{-4.4}$ & $5.2^{+3.4}_{-2.9}$  & $4.615^{+0.059}_{-0.180}$  & $10.78^{+0.14}_{-0.42}$ & $3.88^{+0.09}_{-0.17}$               & $4.98^{+0.31}_{-0.39}$ & $11.63^{+0.71}_{-0.90}$ & $4.19^{+0.27}_{-0.33}$               \\
6  & A    & 25106 & Q       & $23.5^{+3.7}_{-3.3}$ & $15.5^{+3.5}_{-3.5}$ & $1.6^{+2.6}_{-1.6}$  & $3.079^{+0.098}_{-0.146}$  & $7.19^{+0.23}_{-0.34}$  & $2.59^{+0.10}_{-0.13}$               & $3.37^{+0.26}_{-0.30}$ & $7.88^{+0.61}_{-0.69}$  & $2.84^{+0.22}_{-0.25}$               \\
7  & A    & 25106 & F       & $25.4^{+5.2}_{-4.6}$ & $30^{+17}_{-15}$     & $46^{+14}_{-10}$     & $7.65^{+0.16}_{-0.50}$     & $17.86^{+0.37}_{-1.16}$ & $6.44^{+0.18}_{-0.43}$               & $6.15^{+0.50}_{-0.51}$ & $14.4^{+1.2}_{-1.2}$    & $5.18^{+0.43}_{-0.44}$               \\
8  & A    & 25106 & E       & $28.4^{+4.8}_{-4.2}$ & $20.9^{+5.6}_{-7.7}$ & $11.9^{+6.1}_{-5.0}$ & $4.65^{+0.12}_{-0.26}$     & $10.86^{+0.27}_{-0.62}$ & $3.91^{+0.12}_{-0.23}$               & $4.60^{+0.35}_{-0.40}$ & $10.74^{+0.82}_{-0.94}$ & $3.87^{+0.30}_{-0.34}$               \\
9  & B    & 25103 & Q       & $11.7^{+2.0}_{-1.8}$ & $12.9^{+1.9}_{-1.7}$ & 0$^\dagger$          & $1.773^{+0.047}_{-0.072}$  & $4.13^{+0.11}_{-0.17}$  & $3.30^{+0.14}_{-0.17}$               & $1.97^{+0.14}_{-0.17}$ & $4.58^{+0.33}_{-0.39}$  & $3.66^{+0.29}_{-0.33}$               \\
10 & B    & 25104 & Q       & $16.5^{+2.5}_{-2.2}$ & $12.8^{+3.7}_{-3.8}$ & $10.0^{+3.0}_{-3.1}$ & $2.819^{+0.053}_{-0.109}$  & $6.57^{+0.12}_{-0.26}$  & $5.25^{+0.19}_{-0.26}$               & $2.91^{+0.19}_{-0.23}$ & $6.78^{+0.44}_{-0.54}$  & $5.42^{+0.39}_{-0.46}$               \\
11 & B    & 25105 & Q       & $14.5^{+2.4}_{-2.1}$ & $11.0^{+3.6}_{-3.6}$ & $7.6^{+2.7}_{-2.9}$  & $2.386^{+0.059}_{-0.102}$  & $5.56^{+0.14}_{-0.24}$  & $4.44^{+0.18}_{-0.23}$               & $2.49^{+0.17}_{-0.21}$ & $5.80^{+0.40}_{-0.48}$  & $4.63^{+0.35}_{-0.41}$               \\
12 & B    & 25105 & E       & $22.4^{+4.3}_{-3.7}$ & 0$^\dagger$          & $53.8^{+4.6}_{-4.5}$ & $5.36^{+0.16}_{-0.26}$     & $12.50^{+0.38}_{-0.60}$ & $9.99^{+0.44}_{-0.56}$               & $4.77^{+0.37}_{-0.42}$ & $11.11^{+0.87}_{-0.98}$ & $8.88^{+0.75}_{-0.82}$               \\
13 & B    & 25106 & Q       & $18.3^{+2.7}_{-2.4}$ & $4.4^{+3.5}_{-3.6}$  & $10.6^{+2.7}_{-2.6}$ & $2.526^{+0.064}_{-0.107}$  & $5.89^{+0.15}_{-0.25}$  & $4.71^{+0.19}_{-0.24}$               & $2.62^{+0.19}_{-0.22}$ & $6.11^{+0.45}_{-0.51}$  & $4.88^{+0.39}_{-0.43}$               \\ \hline
\end{tabular}
\tablecomments{
    $a$: $10^{51}$\,cm$^{-3}$.
    $b$: $10^{-12}$\,erg\,s$^{-1}$\,cm$^{-2}$.
    $c$: $10^{29}$\,erg\,s$^{-1}$.
    $^\dagger$: Parameter best fit value was negligible in the initial fit and the fit was redone with parameter fixed to 0.
    The stellar states are quiescent (Q), elevated (E), or flare (F), as per the temporal definitions displayed in Fig. \ref{fig:xLC}. We give the unabsorbed fluxes at Earth, luminosities, and ratios to the bolometric luminosity in two different bands: the observed 0.5--10\,keV band, and 0.1--2.4\,keV, which allows for direct comparison with the \citet{Wright2011,Wright2018} sample.}
\end{table*}

We extracted 13 separate spectra across the four observations and two stars: eight for star A, and five for star B. These correspond to one for each star's quiescent spectrum in each observation, plus one for each flare and period of elevated count rate (see definition in Section \ref{ssec:xLC} and Fig. \ref{fig:xLC}). Each spectrum, together its best fit model (discussed below), is plotted in Fig.~\ref{fig:xSpec}.

Visual examination of the spectra revealed them to be dominated by flux between 0.8 and 2 keV. Significant emission at these wavelengths, especially just below 1\,keV, is typical of young stars \citep[e.g.][]{Guedel1997}. We note several broad features in the spectrum. The first peaks at around 1\,keV, and is likely associated with Fe L shell emission, although O and Ne also have relatively strong lines at these energies. Bumps at around 1.4 and 1.9\,keV are also visible in some of the spectra, and are likely associated with Mg (XI and XII) and Si (XIII and XIV), respectively. Lines associated with all of these elements except Si were previously resolved in \textit{XMM-Newton} observations with the RGS, as reported by \citet{Pillitteri2022}.

We fitted all 13 spectra simultaneously in \texttt{PyXspec} (using \texttt{Xspec} version 12.11.1; \citealt{Arnaud1996}) with \texttt{apec} models, which describe emission from a collisionally-ionized plasma \citep{Smith2001}. We accounted for interstellar absorption using a \texttt{tbabs} model \citep{Wilms2000} with the hydrogen column density fixed to $10^{20}$\,cm$^{-2}$, in line with the values in Table 2 of \citet{Pillitteri2022}. However, reducing this value by up to an order of magnitude had a negligible effect on our fitted parameters, likely because our observations contained relatively little flux below 0.5\,keV, where the effects of absorption by the interstellar medium are at their strongest. In all of our fits, we set the Solar abundances to the values determined by \citep{Asplund2009}, and used Cash statistics in order to find the best model parameters \citep{Cash1979}.

Our fits employed three model temperatures for each star, linked across all spectra for that star. The normalizations, and thus emission measures, associated with these temperatures were allowed to vary. Our initial models where we used either fixed Solar abundances, or a single scaling factor to Solar for all elements both yielded poor fits. Therefore, we used the \texttt{vapec} variant, which permits the abundance of individual elements to be changed. We freed up the abundances of O, Ne, Mg, Si, and Fe, but forced them to be the same across all spectra. The choice of these five elements were motivated by the observed spectral features, described above. This fit yielded a good fit to the data (p=0.53).

The best fit temperatures and abundances are given in Table \ref{tab:temp+abund}, while the best fit emission measures and fluxes are displayed in Table \ref{tab:EM+fluxes}, together with the corresponding X-ray luminosity and ratio of this to the bolometric luminosity. We list unabsorbed fluxes and luminosities for two different energy bands. The first of these is the ``observed'' 0.5 -- 10\,keV band, with fluxes and luminosities associated with the band denoted as $F_{\rm x, 0.5}$ and $L_{\rm x, 0.5}$, respectively. The second band we provide fluxes and luminosities for is the 0.1 -- 2.4\,keV band, denoted by $F_{\rm x, 0.1}$ and $L_{\rm x, 0.1}$. This is the ROSAT energy band, which allows direct comparison with empirical X-ray relations (see Section \ref{sec:compRel}). 

The effective area of the \textit{Chandra} ACIS-S instrument reduces significantly below 0.5\,keV, making it difficult to obtain estimates for the flux down to 0.1\,keV. Comparing our spectra with those from EPIC-MOS in \citet{Benatti2021} reveals that our observations drop in counts more sharply below 0.5\,keV, and as such our best fitted model does not contain a component at 0.3\,keV, in contrast to their best fit model. The ACIS-S response files are also only calculated down to 0.3\,keV. We checked which of our spectra had a best fit flux that most closely matched that of \citet{Benatti2021} for \DSTA\ in the 0.5 -- 10\,keV band (spectrum 6, for the quiescent state in 25106). We then compared our 0.3 -- 2.4\,keV flux for that spectrum to theirs in the 0.1 -- 2.4\,keV band, finding the latter to be about 10\% higher. We therefore estimate $F_{\rm x, 0.1}$ 
by scaling the 0.3-2.4 keV flux up by 10\%, and we also added in an extra 5\% fractional uncertainty in each direction to account for this process. We emphasize that the most dominant component in the \citet{Benatti2021} model a plasma at 0.95\,keV, a value in line with the considerable emission we also observe in this region, and that the age of the system means the ``missing" flux at soft energies due to the energy range covered by ACIS-S is relatively small. Were the DS\,Tuc system a few Gyr old, the X-ray spectrum would likely be substantially softer and dominated by energies between 0.1 and 0.5\,keV, and the process of estimating $F_{\rm x, 0.1}$ 
we used here would introduce far more uncertainty.

\section{\textit{Swift} observations and results}
\label{sec:Swift}

\begin{figure}
\centering
 \includegraphics[width=\columnwidth]{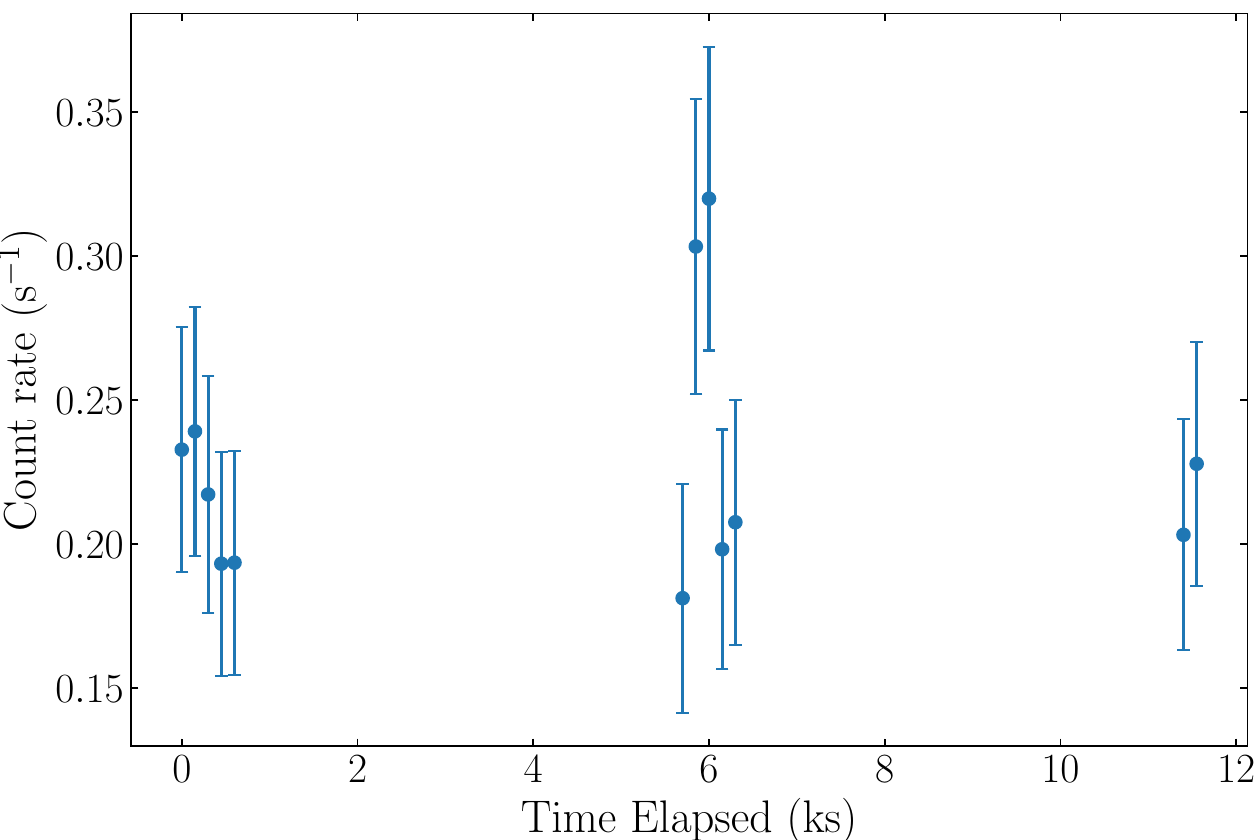}
 \caption{Swift XRT light curve of DS\,Tuc in the 0.3--2.4\,keV energy band.}
 \label{fig:swiftLC}
\end{figure}

\begin{table}
\label{tab:Swift}
\centering
\caption{Best-fit parameters and fluxes from our \textit{Swift} XRT spectral analysis.}
\begin{tabular}{lcc}
\hline
Parameter       & Value                                               & Unit                     \\ \hline
$kT_{\rm 1}$    & $0.809^{+0.073}_{-0.093}$                           & keV                      \\
$kT_{\rm 2}$    & $1.54^{+1.18}_{-0.31}$                              & keV                      \\
EM$_{\rm 1}$    & $\left(3.00^{+0.73}_{-0.84}\right) \times 10^{52}$  & cm$^{-3}$                \\
EM$_{\rm 2}$    & $\left(2.55^{+0.79}_{-0.83}\right) \times 10^{52}$  & cm$^{-3}$                \\
$F_{\rm x,0.3}$ & $\left(4.06^{+0.38}_{-0.13}\right) \times 10^{-12}$ & erg\,s$^{-1}$\,cm$^{-2}$ \\
$F_{\rm x,0.1}$ & $\left(4.47^{+0.59}_{-0.33}\right) \times 10^{-12}$ & erg\,s$^{-1}$\,cm$^{-2}$ \\ \hline
\end{tabular}
\tablecomments{The unabsorbed fluxes at Earth, $F_{\rm x,0.3}$ and $F_{\rm x,0.1}$, correspond to the 0.3--2.4\,keV and 0.1--2.4\,keV energy bands, respectively. The values given here are for a source region encompassing both stars, due to the lower spatial resolution of \textit{Swift} XRT as compared to \textit{Chandra}.}
\end{table}

In addition to the \textit{Chandra} observations presented in this work, and the previous publications of the \textit{XMM-Newton} observations, the DS\,Tuc system was also observed for 2.7\,ks with \textit{Swift} XRT on 2020 Mar 16, spread across three orbits of the spacecraft. 

We originally analyzed this observation as part of a wider analysis of \textit{Swift} data of planetary systems discovered by \textit{TESS}, to be published in an separate paper (Hernandez et al., in prep). For this work, we reanalyzed the XRT spectrum using \texttt{Xspec} version 12.11.1, fixing the abundances to those we found in Section \ref{ssec:xSpec}. Unfortunately, the spatial resolution of the XRT is insufficient to resolve the two separate stars in the DS\,Tuc system, as demonstrated in Figure \ref{fig:imageComp}, and we therefore used a single source region encompassing both stars. Given the lower number of counts compared to \textit{Chandra}, we used only two \texttt{vapec} models, instead of three.

The results of our spectral analysis are in Table \ref{tab:Swift}. The best-fit temperatures from this analysis are consistent with the two lowest temperatures from the Chandra analyses for both stars within 1-$\sigma$. We derived $F_{\rm x,0.1}$ from the 0.3--2.4\,keV fluxes in the same way as in the \textit{Chandra} analysis in Section \ref{ssec:xSpec}. The \textit{Swift} $F_{\rm x,0.1}$ is lower than the equivalent combined quiescent fluxes for stars A and B in all four of the \textit{Chandra} observations. The significance of the difference varies between 2.2-$\sigma$ lower when compared to \textit{Chandra} observation 25106 and 5.4-$\sigma$ lower when compared to observation 25104. These $\sigma$ differences may also be slightly underestimated because, as we will now discuss, we cannot be confident that both stars were in a quiescent state during the \textit{Swift} observation.

Fig.~\ref{fig:swiftLC} shows the 0.3--2.4\,keV light curve of the 40$"$ source region we employed in our analysis at 150\,s cadence, chosen to ensure a few tens of counts in each time bin. The light curve reveals a possible flare during the second of the three \textit{Swift} snapshots. The second and third time bins from the second orbit show an approximately 50\% increase, as compared to the rest of the points. However, within the relatively large uncertainties, there was no clear increase in the hardness ratio during the rise in count rate, as we see for the \textit{Chandra} flares. Additionally, the spatial resolution of \textit{Swift} XRT is insufficient to place constraints on which of the two stars the increased count rate originated from. We conclude that while there is some evidence of flaring, we cannot unambiguously determine if the rise was due to a bona fide stellar X-ray flare from one of the two stars.

\section{Flares and their contribution to the irradiation of the planet}
\label{sec:flares}

In section \ref{ssec:xLC}, we identified two flares from \DSTA\ in our \textit{Chandra} observations, to go with two flares previously identified in \textit{XMM-Newton} observations. We also identified two further periods of elevated count rate, above defined periods of quiescence. Here, we now contextualize these flares, elevated, and quiescent periods, and determine their relative contributions to the overall X-ray emission of \DSTA. We do not discuss the possible \textit{Swift} flare in this section due to the various ambiguities surrounding it, and also focus our attention only on the planet hosting star A.

\subsection{Flaring rates and energetics}

In order to calculate various flaring rates across the \textit{Chandra} and \textit{XMM-Newton} data, we compare the frequency of flares to the elapsed time on target across the observations with each facility. This is as opposed to using the live time, because the temporal resolution of X-ray variability we are sensitive to is much longer than the dead spaces between live time periods\footnote{Live time is the total exposure time, for which the \textit{Chandra} ACIS definition is discussed here: \url{https://cxc.cfa.harvard.edu/ciao/ahelp/times.html}}.

The total elapsed time for \textit{Chandra} and \textit{XMM-Newton} is 63.25\,ks and 65.42\,ks, respectively. Each detected two flares of star A across their observations, leading to very similar flaring rates of 2.7 and 2.6 per day. The combined rate across all observations with both telescopes is 2.7 X-ray flares per day.

We also estimated the energy of the two \textit{Chandra} flares, by taking 
the luminosities for those periods in Table \ref{tab:EM+fluxes} to be representative of the average luminosity across the flare, and multiplying by the duration. We subtract off the contribution by quiescent emission to the flaring epoch, which we assume is the luminosity of the measured quiescent period elsewhere in the same observation. 
This method also assumes all emission, quiescent or otherwise, to be directionally homogenous. We calculated the total energy emitted by the flares in the 0.5--10\,keV band in observations 25104 and 25106 to be $8.6\times10^{33}$\,erg and $3.7\times10^{33}$\,erg, respectively.

Our calculated flare energies are about an order of magnitude less than the reported energies in the 0.3--10\,keV band of the two \textit{XMM-Newton} flares \citep{Pillitteri2022}. The relative dimness of the \textit{Chandra} flares compared to \textit{XMM-Newton} makes sense when comparing our light curves in Fig \ref{fig:xLC} to their fig. 2. The \textit{XMM-Newton} flares show a larger factor increase in the peak count rate compared to quiescent, as well as a longer duration. \citet{Dethero2023} also detected a powerful (($2.8\pm0.1)\times10^{35}$\,erg) flare in the DS\,Tuc system using \textit{NICER}, which they attributed to star A. However, we note that \textit{NICER} lacks the spatial resolution to distinguish the binary, and so it is just as likely that the flare they observed was from star B as it is star A.

To put the DS\,Tuc flares into a wider context, we note that soft X-ray flares of similar energies have been measured for range of Solar and later type stars \citep[e.g.][]{Pye2015}, with a few flares reaching as high as $\sim10^{37}$\,erg \citep{Zhao2024}. They are more energetic than the for the Sun, though, with the most energetic soft X-ray Solar flares measured with the Geostationary Operational Environmental Satellite system reaching integrated energies on the order of $10^{31}$\,erg \citep{Plutino2023}. This is in a more restrictive energy band (0.15--1.24\,keV), but highlights the relative weakness of Solar flares compared to many other stars in the Galaxy.

\citet{Colombo2022} previously assessed flare frequencies in optical light, based on \textit{TESS} light curves, finding a similar flare rate of two per day for optical flares above $2\times10^{32}$\,erg. However, \DSTA\ and B fit within one TESS pixel, and so this is very likely a rate for both stars combined. The \citet{Colombo2022} study also predicted two X-ray flares per day with energies of $2\times10^{31}$\,erg, based on a relation derived by \citet{Flaccomio2018}. The X-ray data across \textit{Chandra} and \textit{XMM-Newton} appear to agree with this prediction, however, we note that the four unambiguously detected X-ray flare events so far have energies well in excess of this value, by up to three orders of magnitude. \citet{Pillitteri2022} interpret those detected by \textit{XMM-Newton} as very rare events due to their large detected energies and the implication for their optical counterparts ($\gtrsim 5 \times 10^{35}$\,erg). While the \textit{Chandra} flares we present here are an order of magnitude lower in X-ray brightness, together with the \textit{XMM-Newton} data, they demonstrate that X-ray flares from \DSTA\ emitting well above the $2\times10^{31}$\,erg value suggested by \citet{Colombo2022} may be a common occurrence. A possible reason for this difference could be that the \citet{Flaccomio2018} relation they used was based solely on a sample of 3\,Myr old pre-main-sequence stars. Such stars are still spinning up and so the relation may not be applicable to a 45\,Myr star like \DSTA.

\subsection{Relative quiescent and flaring contributions}

We calculated the relative contributions of the three types of defined epoch (quiescent, elevated, and flare), both temporally and energetically. Across the total 63.25\,ks of elapsed time, epochs we defined as quiescent account for 66.1\%. Flaring periods make up a further 17.3\%, with the remaining 16.5\% defined as elevated\footnote{numbers add to 99.9\% due to rounding}.

By applying the same constant quiescent subtraction assumptions to the elevated epochs, we then calculated the relative energetic contributions of each emission type. We find that quiescent emission is responsible for 78\% of the total X-ray emission across the \textit{Chandra} observations. This is higher than the percentage time in the quiescent epoch because of our assumption that the quiescent contribution continues unchanged during flaring and elevated periods. We determine 15\% of observed emission in \textit{Chandra} is from flares. The elevated epochs account for the remaining 7\% of the observed emission. The larger flares in the \textit{XMM-Newton} data suggest the true time-averaged flaring contribution may be somewhat higher than this, but the small numbers of flares detected so far make it difficult to determine.

These assessments demonstrate that despite the large stellar flares that we detected, the majority of the emission we observed can be considered quiescent. As such, when we evaluate the effect of XUV on \DSTA\,b in Section \ref{sec:planetEff}, we use the average measured quiescent flux to inform our calculations. If further observations are made in the future that significantly improve our constraints on the flaring rate of \DSTA, then assessing the contribution of flares to the mass loss of the planet would be worthwhile and an important extension of this work.

\section{Long-term variation}
\label{sec:temporal}

\begin{figure*}
\centering
 \includegraphics[width=\textwidth]{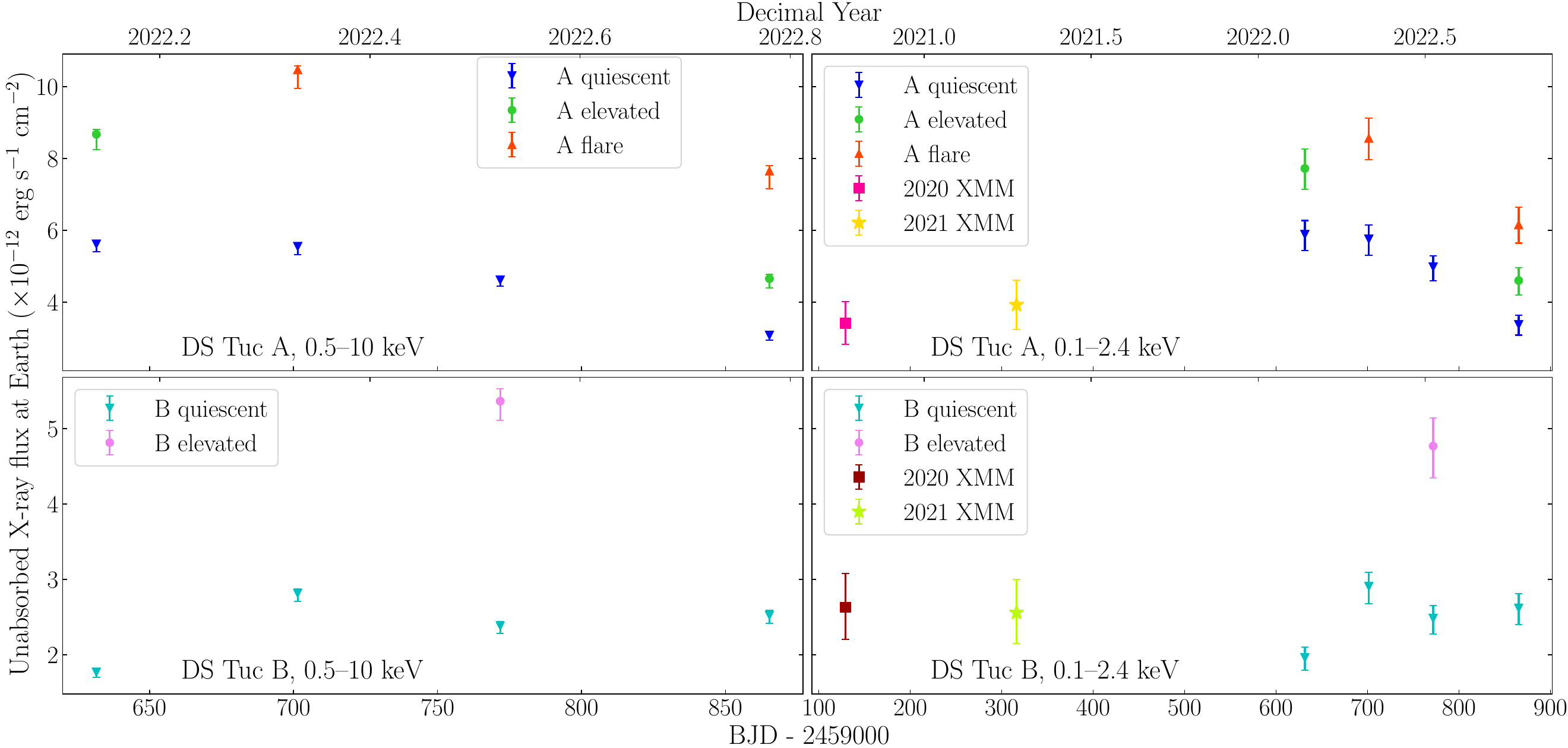}
 \caption{X-ray light curves of \DSTA\ (top panels) and B (bottom panels) across the observations. The left-hand panels are the unabsorbed fluxes at Earth in the observed 0.5--10\,keV band for the \textit{Chandra} observations, plotting quiescent, elevated, and flaring epochs separately. The right-hand panels are the equivalent fluxes extrapolated to the 0.1--2.4\,keV band (as described in \ref{ssec:xSpec}). On the right-hand panels, we also plot for comparison the \textit{XMM-Newton} points from \citet{Benatti2021} and \citet{Pillitteri2022} for the 2020 and 2021 observations, respectively. Colors and symbols for the \textit{Chandra} points are the same as in previous figures.}
 \label{fig:tempVarComb}
\end{figure*}



We assessed the temporal variation of \DSTA\ and B. In the left-hand panels of Fig.~\ref{fig:tempVarComb}, we plot $F_{\rm x,0.5}$ across the \textit{Chandra} observations as a function of observation date, displaying the elevated and flaring periods separately. The upper errorbars of several points are smaller than the points themselves. 

We also plot $F_{\rm x,0.1}$ values from our \textit{Chandra} analysis in the right-hand panels of Fig.~\ref{fig:tempVarComb}. This energy band allows us to also include and compare to the points from \textit{XMM-Newton}. \citet{Benatti2021} provides values of $F_{\rm x,0.1}$ 
for both stars, and we plot these as the ``2020 XMM" points on the right side of Fig.~\ref{fig:tempVarComb}. We note that the errorbars given by \citet{Benatti2021} do not appear to take systematic uncertainty due to source confusion into account. Elsewhere in the paper they state cross-contamination is on the order of 16\%, and so we add a 16\% error in quadrature to their stated statistical errorbars when plotting these points. \citet{Pillitteri2022} does not explicitly state values of the quiescent flux of either star for the 2021 observation, but section 3.1 of their paper does give descriptions of the values as compared to the 2020 observation, together with the estimated contributions of stars A and B. 
From this, we calculated the values plotted on the right side of Fig. \ref{fig:tempVarComb} as the ``2021 XMM" points, and set the uncertainties to the same percentage precision as for the 2020 \textit{XMM-Newton} points. However, the disadvantage of the right-hand panels versus the left-hand panels is that the extrapolation to 0.1\,keV inflates the uncertainties of the \textit{Chandra} points as compared to using the observed bands. The best assessment of the relative precision of the two telescopes for observing DS\,Tuc is therefore made by comparing the \textit{Chandra} errorbars on the left-hand side of Fig.~\ref{fig:tempVarComb} with those for \textit{XMM-Newton} on the right. Accounting for source confusion with the latter means \textit{Chandra} is the clear winner for performing the kind of high precision measurements required for assessing long term variability for this target.

The quiescent points for star A from our 2022 \textit{Chandra} observations show a clear, monotonic decline in the measured flux across the four observations, with the first observation being 1.8 times brighter than the last. The decline is such that the flaring epoch in the 25106 observation has an average flux only moderately higher than the quiescent level in 25103 and 25104. The comparison \textit{XMM-Newton} points both lie within the range of the \textit{Chandra} measurements, suggesting the amplitude of quiescent variation of the star is perhaps not much larger than this 1.8 factor among the \textit{Chandra} observations.

The origin of the observed decline in the \textit{Chandra} data is not certain, but one possibility is they hint at a Solar-like activity cycle.  \citet{Savanov2020} analyzed visual magnitude data for DS\,Tuc from the All Sky Automated Survey whose power spectrum revealed possible cyclical behavior with periods of 4.4\,yr and 360-400\,d, though they attribute the latter to the seasonal nature of the ground-based observations. We note however that \citet{Basri2020} showed that short period ``cycles" identified in optical data can be the result of random processes.

The existing data are insufficient to examine any periodicity of the X-ray signal. However the decline in the \textit{Chandra} points across almost eight months suggests any associated periodicity would likely be at least twice as long, in line with the possible optical signal. Among stars with measured coronal cycles there appears to be a relationship between cycle amplitude and Rossby number (see fig. 9 of \citealt{Wargelin2017} and \citealt{Coffaro2022}). In Section \ref{ssec:X-Ross}, we estimate the Rossby number of \DSTA\ to be $0.175^{+0.020}_{-0.016}$, a value for which an activity cycle amplitude would be expected to be much smaller than the measured 1.8 factor decline, based on this observed relationship. However, \DSTA\ is an order of magnitude younger than the youngest star for which a coronal activity cycle has been confirmed \citet{Coffaro2022}, and so the validity of this relationship at young ages, and therefore small Rossby numbers, is unknown. We note that another young planet host, V1298\,Tau, also shows variability on a similar level to \DSTA, but again the cyclicity of this signal is unknown \citep{Maggio2023}.
 
The observed decline demonstrates \DSTA\ is an exciting target for understanding the X-ray variation of young Solar-like stars on timescales of months to years.  More \textit{Chandra} observations of DS\,Tuc taken over the next few years would enable a greater understanding of the origin of this decline, and be able to determine if it is part of a periodic signal, along with its characteristics if so.




The quiescent points for star B exhibit some variation between epochs, but with no clear overall trend. Most notable is the anomalously low flux in the first observation, which is a factor 1.6 lower than the highest quiescent flux. Even if we exclude this low point, the other three show variability that is significant given the tight uncertainties on the points, albeit with a smaller amplitude, remaining within 10\% of their mean. The \textit{XMM-Newton} points for star B are broadly consistent with the range covered by the \textit{Chandra} points within the relatively large uncertainties of the \textit{XMM-Newton} points. 

\textit{ROSAT} and \textit{eROSITA} have also both detected X-ray emission from the region of DS\,Tuc. Both telescopes lack the spatial resolution to separate the two binary components, and so they cannot provide further information of the temporal variation of the two individual stars. In terms of the total luminosity across both stars, \textit{ROSAT} measured $2.4 \times 10^{30}$\,erg\,s$^{-1}$ in the 0.1--2.4\,keV band \citep{Benatti2021}. The average A+B total luminosity in the same band across the quiescent \textit{Chandra} measurements is $1.75 \times 10^{30}$\,erg\,s$^{-1}$. \citet{Foster2022} reported an \textit{eROSITA} luminosity for \DSTA\, but this assumes a 50-50 contribution of the two stars. Our \textit{Chandra} results demonstrate this ratio is not equal and somewhat time varying. Doubling the \citet{Foster2022} value to obtain a total A+B luminosity yields $1.43 \times 10^{30}$\,erg\,s$^{-1}$ in the slightly more restrictive 0.2--2.0\,keV band. The results from all four telescopes together show that long term variation in the total luminosity of the two stars larger than a factor of a few is unlikely.

\section{Comparison with empirical X-ray relations}
\label{sec:compRel}

We compared the 0.1 -- 2.4\,keV fluxes calculated in Section \ref{ssec:xSpec} to empirical relationships between X-ray emission and both stellar rotation period and age. Note that the \textit{XMM-Newton} points were not included in the analysis in this section due to their much larger uncertainties, and to ensure the points used for our statistical calculations were obtained in a self-consistent manner. The advantage of including them in Section \ref{sec:temporal}, i.e. elongating the baseline for assessing the temporal variation, is also not relevant here.

\subsection{Comparison to the X-ray-Rossby number relationship}
\label{ssec:X-Ross}

\begin{figure}
\centering
 \includegraphics[width=\columnwidth]{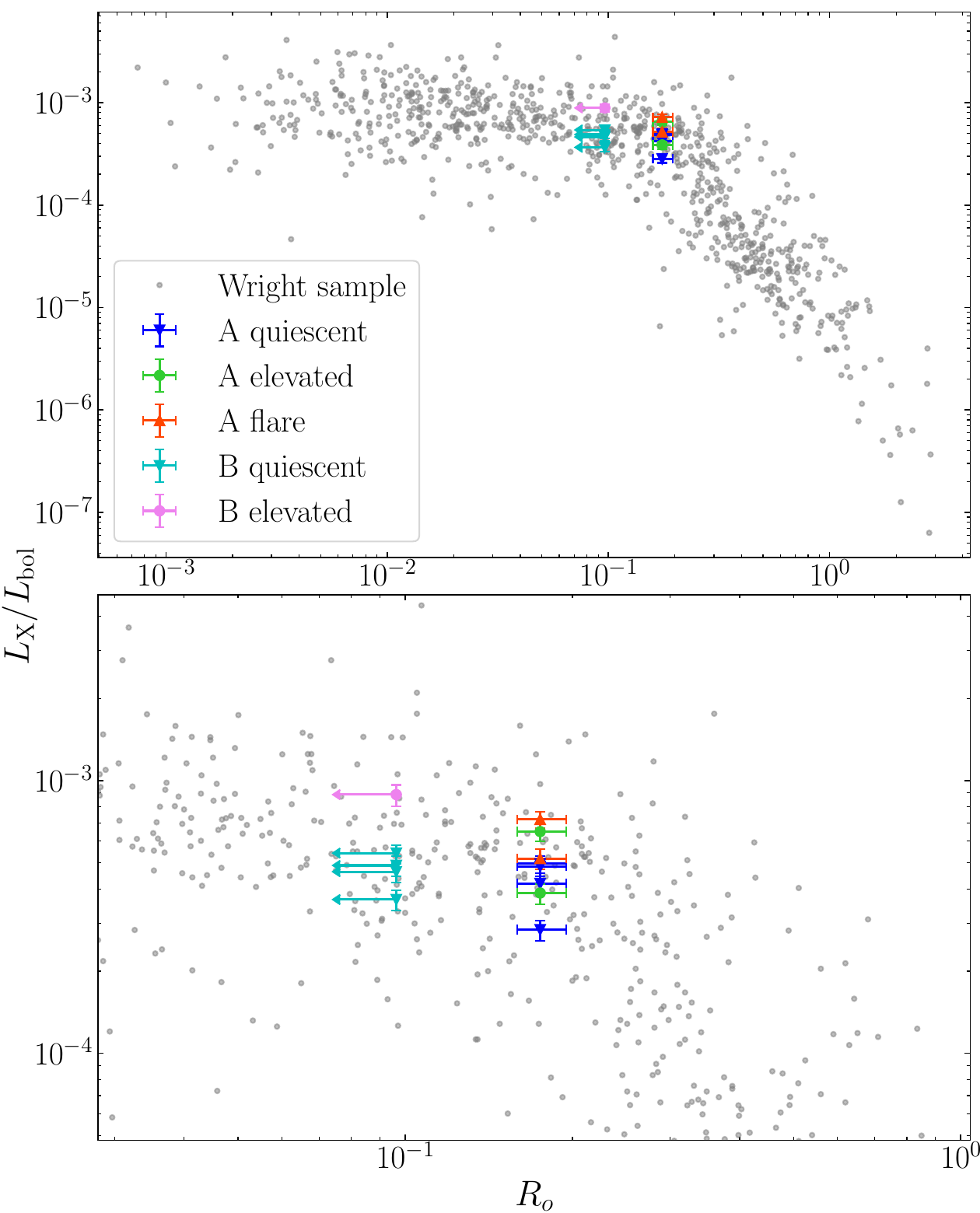}
 \caption{The \citet{Wright2018} sample with our measurements of \DSTA\ and B overplotted. The bottom panel is a zoom in of the top panel on to the region around where our measurements cluster.}
 \label{fig:wrightComp}
\end{figure}

Our first comparison was to the relationship between $L_{\rm x, 0.1}/L_{\rm bol}$ and the Rossby number, $R_o$, defined as the ratio of the stellar rotation period, $P_{\rm rot}$, and the convective turnover time, $\tau$. For star A, we adopt $P_{\rm rot} = 2.85^{+0.04}_{-0.05}$\,d, as determined by \citet{Newton2019}. Star B however does not have a measured rotation period, and thus we calculate an upper limit on its value using the $v \sin i$ measurement of $14.4\pm0.3$\,km\,s$^{-1}$, also reported by \citet{Newton2019}, yielding $P_{\rm rot} < 3.04$\,d.

To estimate $\tau$, we used the theoretical models of \citet{Landin2023}, which provide values of the convective turnover time based on a discrete set of stellar masses. The closest age to DS\,Tuc for which values were provided on the 1\,M$_\odot$ track was 32.8\,Myr, and led us to adopt $\tau = 16$\,d for star A. The 0.8\,M$_\odot$ track provides values at different ages, with 44.4\,Myr being closest to DS\,Tuc. Our resulting adopted value of $\tau$ for star B was 31\,d. \citet{Landin2023} stated that the difference in $\tau$ between different versions of their model was 5\%, but we adopt more conservative uncertainties on $\tau$ of 10\% to account for e.g. the difference in age between the actual system and the closest point on the model track for which $\tau$ is given. Therefore, we estimate the Rossby number of the star A to be $0.175^{+0.020}_{-0.016}$, and an upper limit for star B of $<0.096$. These values lie just either side of the empirical $R_o = 0.13$ determined by \citet{Wright2011,Wright2018} as the boundary between the saturated and unsaturated regimes.

\begin{table}
\centering
\caption{Statistical tests for the DS\,Tuc stars and comparison samples.}
\label{tab:variationComp}
\begin{tabular}{lccc}
\hline
Sample       & St. Dev          & Range            & KS-Test               \\
             & $\times 10^{-4}$ & $\times 10^{-4}$ & p-value               \\ \hline
Star A       & 1.31             & 4.37             & \multirow{2}{*}{0.52} \\
Comparison A & 3.09             & 15.82            &                       \\ \hline
Star B       & 1.79             & 5.22             & \multirow{2}{*}{0.44} \\
Comparison B & 4.13             & 26.88            &                       \\ \hline
\end{tabular}
\tablecomments{These are the standard deviation and range of our $L_{\rm x, 0.1}/L_{\rm bol}$ measurements for each star. We compare these to samples of the closest 100 in R$_o$ to each star in the \citet{Wright2018} sample. In the final column we give the p-value from a KS-test of the two samples.}
\end{table}

The series of papers by \citet{Wright2011,Wright2016,Wright2018} built a sample of over 800 stars ranging in type from F through M with $L_{\rm x, 0.1}/L_{\rm bol}$ and R$_o$ values. In Fig.~\ref{fig:wrightComp}, we plot our values for DS\,Tuc\,A and B, together with the sample from \citet{Wright2018}. The sample shows scatter of over an order of magnitude in $L_{\rm x, 0.1}/L_{\rm bol}$ at a given value of R$_o$. In comparison, our points seem to trace out a considerable portion of the scatter, particularly for star B.

To be quantitative, in Table \ref{tab:variationComp}, we display the standard deviation and range of various samples of points: our measurements for star A, our measurements for star B, and comparison subsamples of the \citet{Wright2018} sample. For star A, the comparison subsample was made up of the 100 stars closest in R$_o$. For star B, we used the 100 stars closest to its R$_o$ upper limit whose value did not exceed it. The ranges and standard deviations for our measurements are typically a factor of a few smaller than the comparison subsamples. However, our relatively low number of measurements for each star are unlikely to trace out the full variation that each star exhibits. We therefore performed a Kolmogorov–Smirnov (K-S) test to examine if the distribution of our measurements matches the subsamples, with these results also given in Table~\ref{tab:variationComp}. Those tests demonstrated that the variation among our measurements and the spread in $L_{\rm x, 0.1}/L_{\rm bol}$ of the stars closest to them in R$_o$ cannot be ruled out from having been drawn from the same distributions (Since $p > 0.05$). Therefore, from these observations, it remains possible that temporal variation in stars is a major contributor to the scatter observed in activity relationships.

Some of our plotted points are for flaring and other elevated periods. While one may be temped to have excluded these points from the tests, there are likely also numerous flaring periods among the measurements used to build the \citet{Wright2018} sample. Both \citet{Wright2011} and \citet{Wright2018} describe efforts to mitigate sample bias due to flares, but particularly among the \textit{ROSAT} All-Sky Survey results it is very likely that some of the sources were detected \textit{only} in flare. We note also that re-performing our K-S tests with only the four quiescent measurements for each star retains the consistency with their respective comparison samples, with p-values of 0.30 and 0.11 for stars A and B, respectively.

\subsection{Comparison to X-ray-age relations}

\begin{figure}
\centering
 \includegraphics[width=\columnwidth]{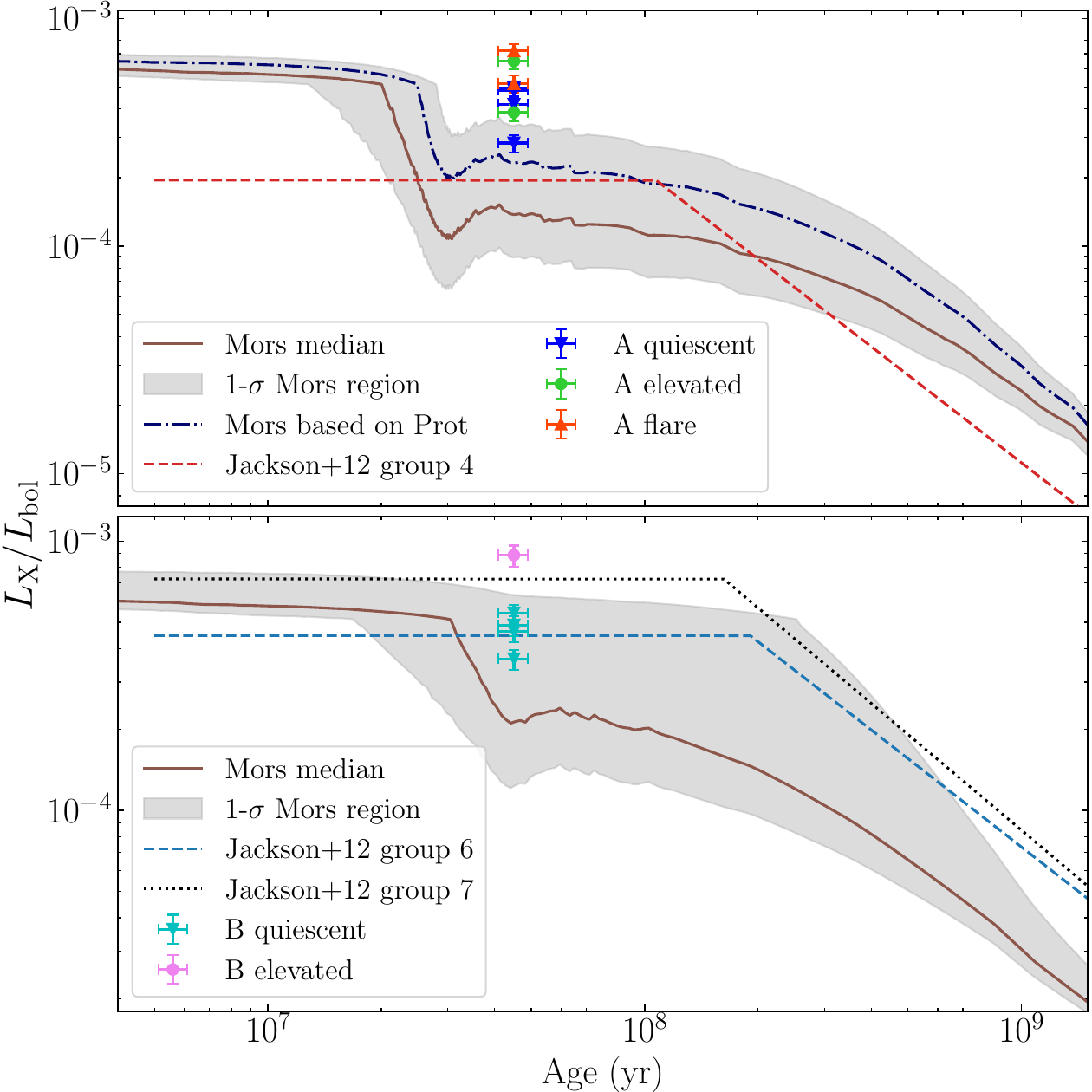}
 \caption{Comparison of our measured values of $L_{\rm x, 0.1}/L_{\rm bol}$ to the X-ray-age relations in \citet{Jackson2012} and evolution profiles generated by the Mors code \citep{Johnstone2021}. Star A is in the top panel, and star B in the bottom panel. For the \citet{Jackson2012} comparisons, the group numbers refer to the $B-V$ color bins in their Table 2, using a system of numbering the bins in ascending order going down the table. We plot two groups for star B, as its $B-V$ value is close to the boundary of two color bins. For the Mors comparisons, we plot the median track, and shade a region around it to show the 1-$\sigma$ confidence region based on the 16th and 84th percentile tracks. For star A, we plot an additional track based on the measured $P_{\rm rot}$. Star B does not have such a track plotted as it lacks a measured $P_{\rm rot}$.}
 \label{fig:xray-age_comb}
\end{figure}



We compared our measurements to two different X-ray-age relationships. The first was the X-ray-age relationships fitted to observational data by \citet{Jackson2012}, and includes a constant level for the saturated regime, followed by a power law decline after some break point in age. The variables describing the tracks are fit separately to seven different $B-V$ color bins that roughly span FGK stellar types. Numbering the bins in ascending order moving down table 2 in \citet{Jackson2012}, star A is in group 4, with $B-V = 0.772$. Star B's $B-V$ of 1.268 places it in group 6, but very close to the boundary with group 7. Given this proximity, in our comparison between the measurements and the relations, plotted in Fig.~\ref{fig:xray-age_comb}, we display the tracks for both groups 6 and 7. Our measurements for star A suggest it may be slightly brighter than expected for a star of its color and age, with all of the fluxes lying above the relation track. The relations do appear to be a better fit for star B however, particularly the group 6 track.

Next we compare the measurements to the evolutionary tracks we generated using the Mors code \citep{Johnstone2021}, with these also plotted in Fig.~\ref{fig:xray-age_comb}. We took the stellar masses for the two stars (1.01 and 0.84\,M$_\odot$, respectively), and generated three temporal X-ray evolutions for each, corresponding to the 16th, 50th, and 84th percentile tracks. For star A, we generated an additional track, based on the known  $P_{\rm rot} = 2.85$\,d at the current age of 45\,Myr, yielding a track which lies roughly halfway between the 50th and 84th percentile evolutions. As with the comparison to \citet{Jackson2012}, star A appears brighter than expected from the Mors tracks, although in this case some of the dimmer measurements are close to the $P_{\rm rot}$ and 84th percentile tracks, and all quiescent points are well within 3-$\sigma$. For star B, the quiescent points all lie between the 16th and 84th percentiles.

Our comparisons to both of these methods of predicting X-ray-age evolution revealed similar results. Star A appears slightly brighter than predicted for its age, which is unexpected given the star is very consistent with the X-ray-Rossby number relation. The \citet{Jackson2012} relations suggest that the star is still in the saturated regime, where the \citet{Wright2018} relations and the Mors code both suggest the star has perhaps just passed into the unsaturated regime at its age and $R_o$. However, the measurements are still above the \citet{Jackson2012} saturation level for a star of this type, which is a factor of a few below the saturation value fitted across all stellar types in \citet{Wright2011,Wright2018}. Despite the better agreement for star B between the measurements and the various relations, the results for star A hint at the slew of empirical relations not being completely consistent with each other across the parameter spaces they consider. This could result from the different samples and methods employed in these studies, as well as the scatter in the measurements. Together with the temporal variation exhibited in our observations, this underlines the need for actual measurements over relying on empirical relationships, 
and carrying out multiple observations that aim to test a variety of activity states of the star(s).

We note that star-planet interactions (SPI) are not expected at a detectable level in this system. 
Based on equation 1 of \citet{Cuntz2000}, the gravitational perturbation of the planet on the star, $\frac{\Delta g}{g}$, is $1.2 \times 10^{-8}$. This is about three orders of magnitude smaller than for a typical hot Jupiter \citep[c.f. fig. 7 of][]{Ilic2022}, for which there is some evidence of tidal SPI at a detectable level in the X-ray emission of a few binary systems \citep{Poppenhaeger2014,Ilic2022}.  

\section{XUV effect on the planet}
\label{sec:planetEff}

We now explore the possible effects of XUV irradiation on the atmosphere of \DSTA\,b. We considered the future evolution of the planet using evolution simulations, as well as investigating the current state of the planet. The latter will be useful in contextualizing the results from attempts to detect outflows from the atmosphere.

\subsection{Simulations of future evolution}
\label{ssec:Sims}

We ran simulations for the evolution of the planet using the stellar evolution code Modules for Experiments in Stellar Astrophysics \citep[MESA,][]{Paxton2011,Paxton2013,Paxton2015,Paxton2018,Paxton2019,Jermyn2023}. With a few exceptions listed below, our methods and inlists used were the same as in \citet{King2024}, which was largely based on earlier work by \citet{Chen2016}, \citet{Malsky2020}, and \citet{Malsky2023}.

\subsubsection{Methods}

For the XUV history in the simulations, we used the median track generated by Mors \citep{Johnstone2021}, and scaled it to the mean of our quiescent measurements across our four observations. We multiplied the median EUV track by the same factor, and irradiate the planet across the XUV using the sum of the X-ray and EUV tracks. The simulations were run from 6\,Myr (typical disk dispersal time) to an age of 9.2\,Gyr, with older ages outside of the calculable limits in Mors for a star of this mass, due to evolution off the main sequence. 

As in \citet{King2024}, we ran simulations across a grid of starting masses (6 -- 20\,M$_\oplus$ inclusive, step size 1.0\,M$_\oplus$) and envelope mass fractions (0.01 -- 0.30, step size 0.01). The range of starting masses we tested was partially motivated by the upper limit placed on the mass of \DSTA\,b of 14.4\,M$_\oplus$ by \citet{Benatti2021}. However, we chose to test up to 20\,M$_\oplus$ because the confidence level of this upper limit was only 68\%.

We used two methods for calculating mass loss rates in our simulations. The first was the energy-limited escape method \citep{Watson1981,Lammer2003,Erkaev2007}, for which we assumed a canonical value for the efficiency, $\eta=0.15$, and an XUV absorption radius equal to the optically-measured planet radius -- i.e., $\beta =1$ using the formalism in equation 2 of \citet{King2024}. The second method we used was an analytic approximation to the hydrodynamic code ATES \citep{Caldiroli2021}, provided in a follow up paper \citep{Caldiroli2022}. We used this approximation as running the full hydrodynamic code across the millions of points at which we evaluated the tracks was too computationally intensive. Energy-limited escape provides a relatively simple benchmark for mass loss rate calculation which has been widely used in the exoplanet literature \citep[e.g.][]{Lammer2003,LDE2012,Foster2022}. Comparing those results with ATES demonstrates how more a complex treatment based on hydrodynamic simulations deviates from the more simple case.

Following the simulation runs, we determined which tracks successfully reproduced the planet's measured radius, $5.70\pm0.17$\,R$_\oplus$ \citep{Newton2019}, to within 1-$\sigma$ at the closest step in the simulations to known age of the system, 45\,Myr. With only the \citet{Benatti2021} upper limit on the planet's mass currently in the literature, we did not consider present-day mass when determining which evolutionary tracks were successful. We note that our simulations are somewhat agnostic to the boil-off process, which can potentially remove a substantial fraction of the accreted envelope immediately after disk dispersion \citep{Owen2016,Rogers2024}. This is not a major concern as we are mostly interested in the planet's future given its young age, and we are only examining tracks which successfully reproduce the current radius, however they got there.

\subsubsection{Results}

\begin{figure}
\centering
 \includegraphics[width=\columnwidth]{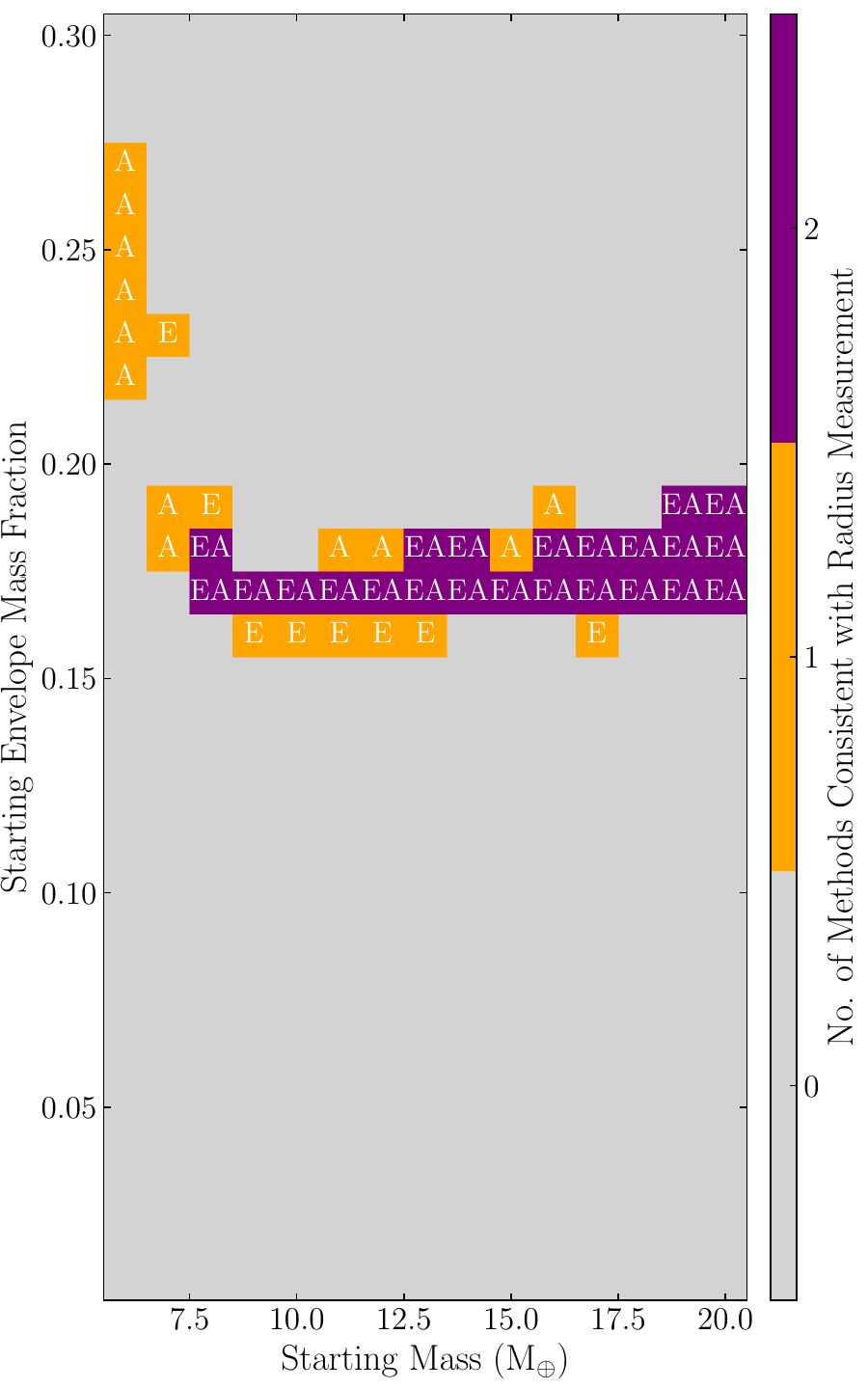}
 \caption{Heatmap depicting the starting grid values in our MESA simulations which were able to reproduce the planet's currently observed radius. The labels show which mass loss methods were successful for that pair of starting values: E - energy-limited escape; A - ATES.}
 \label{fig:heatmap}
\end{figure}

\begin{figure*}
\centering
 \includegraphics[width=0.80\textwidth]{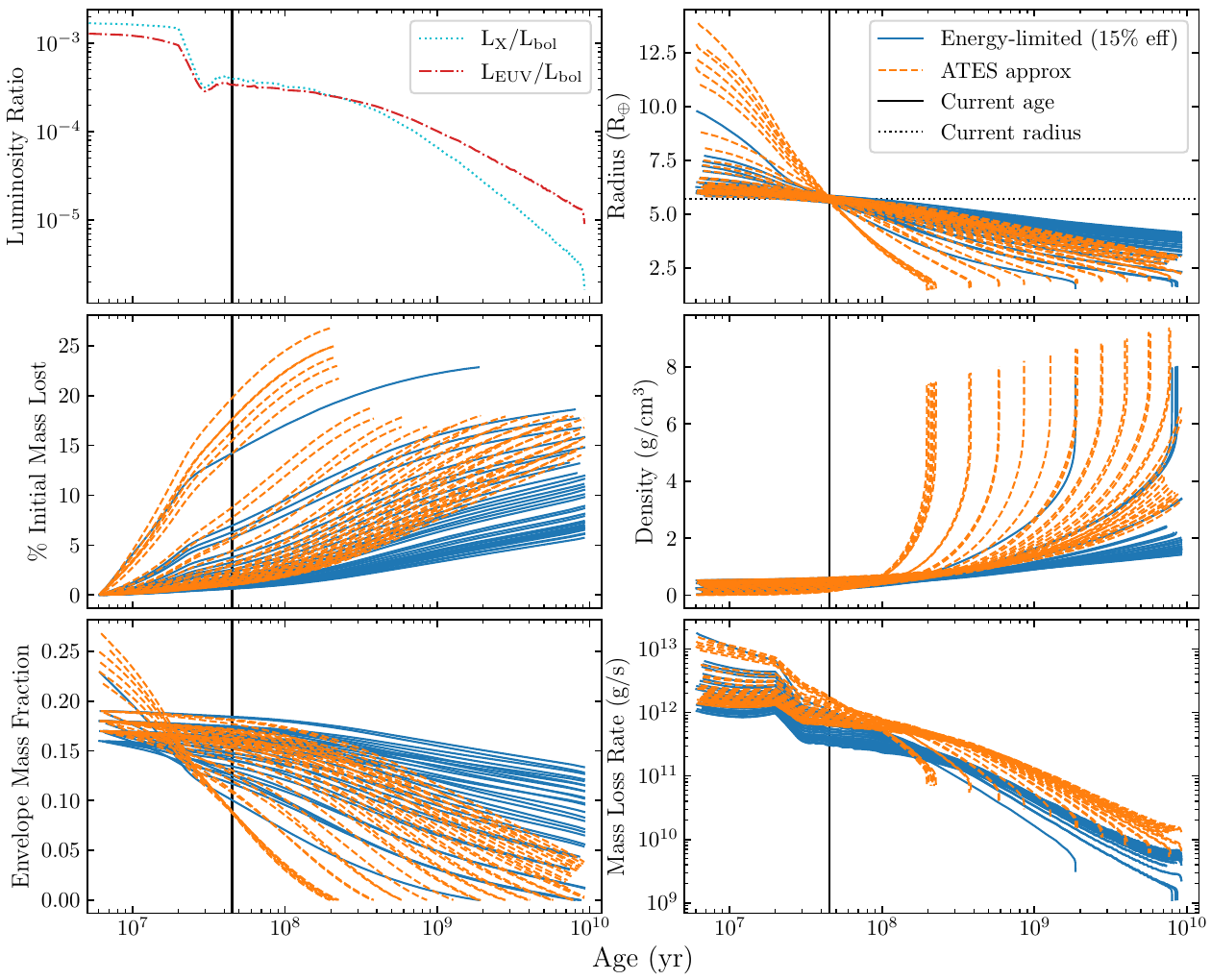}
 \caption{Evolutionary tracks from our simulations which successfully reproduced the measured planetary radius (dotted, horizontal, black line) to within 1-$\sigma$ at the age of the system (solid, vertical, black lines). The top left panel shows the XUV time evolution - specifically their ratios to L$_{\rm bol}$ of the star generated by Mors for \DSTA. The rest of the panels display the time evolution of several key parameters for the planet and the escape of material from it.}
 \label{fig:consistTracks}
\end{figure*}



\begin{figure*}
\centering
 \includegraphics[width=\textwidth]{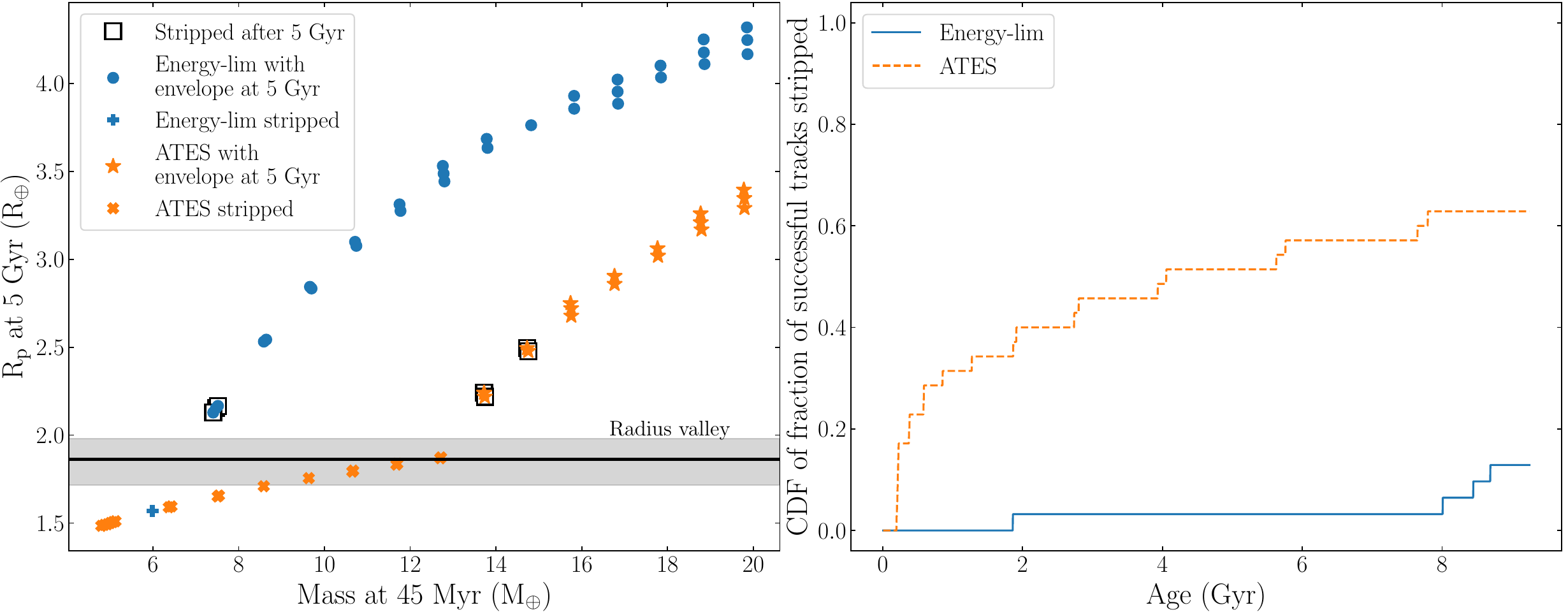}
 \caption{Summary of the outcome of the simulations. The left panel shows the radii of \DSTA\,b at 5\,Gyr as a function of implied current mass for the simulations which reproduced the radius within 1-$\sigma$ at 45\,Myr. We use different symbols to differentiate those that are stripped entirely of their primordial envelope and those which retain at least some portion of it. Those that strip after 5\,Gyr are shown as not stripped here, but highlighted with a box around their symbol. For the tracks reproducing the current observed radius, the right panel shows the cumulative distribution function for the fraction of these successful tracks for which the envelope has been stripped as a function of time. We also mark on the plot the location of the radius valley for a planet with a 8.138\,d orbit like \DSTA\,b (solid black line), together with the 1-$\sigma$ uncertainty region (grey shaded area). These values were calculated using the work of \citet{Ho2023}.}
 \label{fig:5GyrRadii+timescale}
\end{figure*}

Fig.~\ref{fig:heatmap} displays a heatmap of the initial grid, showing which combinations of the starting masses and envelope mass fractions resulted in simulations which reproduced the measured radius. To complement this, Fig.~\ref{fig:consistTracks} shows the evolution of key planetary parameters for the set of successful simulations.

All starting masses we tested had multiple simulation tracks reproduce the measured radius. This is perhaps unsurprising considering we cannot not perform a mass comparison at the current age. However, it also implies that the radius of the planet at this early age is somewhat degenerate with the total mass of the planet, within the range of masses tested. In Fig.~\ref{fig:consistTracks}, most of the radii tracks start closely clustered across both mass loss rate calculation methods and all starting mass values. Only at later age through further mass loss and cooling do the planetary radii of the bulk of successful tracks fan out to a much wider range of values.

There are a few tracks where the radii start much higher and happen to pass through the current radius at 45\,Myr so as to be a plausible solution for \DSTA\,b. These are the few successful tracks where the starting envelope mass fraction is higher than the 0.15-0.20 range exhibited by the majority, as shown in Fig. \ref{fig:heatmap}. Combined with the very low starting masses of these tracks, they are susceptible to losing a substantial portion of their envelope within the first few tens of Myr, and thus shrinking sufficiently to match the measured radii. This also highlights that at these early ages, envelope mass fraction is more important than total mass for setting the radius of the planet.

We also note that across much of the 0.15-0.20 starting range of envelope mass fraction, the two mass loss methods tend to agree on what starting conditions can and cannot produce the correct radius at 45\,Myr. For those that do not agree, the ATES-only successful starting conditions tend to have slightly higher starting envelope mass fractions than those where only energy-limited reproduces the radius. This is most notable among the successful low starting mass, high starting envelope mass fraction tracks, where six of the seven used ATES. Higher possible starting envelope mass fractions for ATES compared to energy-limited makes sense as the mass loss rates are typically higher, as can be seen in the bottom right panel of Figs. \ref{fig:consistTracks}. This is likely because the ATES method allows for $\beta > 1$, in contrast to the fixed value of $\beta = 1$ for the energy-limited calculations.

In the left panel of Fig. \ref{fig:5GyrRadii+timescale}, we plot the planetary radii at a typical field age of 5\,Gyr for all of the tracks which successfully reproduced the measured radius at the current age. These values are plotted against the mass at an age of 45\,Myr in their simulations, for comparison with any future mass measurement. 
For the tracks stripped before 5\,Gyr, we assume there was no further change to their radii after the last point in the MESA track. We also plot the location of the radius valley for a planet with \DSTA\,b's orbital period, based on the empirical relation of \citet{Ho2023}. 

The full sample of tracks show a wide range of radii at 5\,Gyr, from Neptune-size to stripped super-Earth. In general, energy-limited mass loss mostly predicts a Neptune or sub-Neptune-sized future, while ATES predicts a sub-Neptune or stripped super-Earth future. All stripped planets lie in or below the radius valley, and tracks retaining at least a small portion of the primordial envelope are clearly above the radius valley, in line with expectation. All tracks below 8\,M$_\oplus$ at the current age are predicted to lose all of their envelope in the future, while all above 15\,M$_\oplus$ retain at least some portion of it. For those between these mass values, ATES predicts full stripping, while energy-limited does not. A measurement of the current planet mass could make it possible to predict the ultimate fate of the planet, based on these tracks.

Finally, for the simulations where the planet's envelope was stripped, the right-hand panel of Fig. \ref{fig:5GyrRadii+timescale} shows the timescales on which this happened for each mass loss method. In about half of the ATES simulations where the planet is stripped, this happens within about a Gyr. However, there are also tracks where the planet is stripped much later, including well past the typical field age of 5\,Gyr that we used as the test age in the left-hand panel of Fig. \ref{fig:5GyrRadii+timescale}. We highlight such simulations on the left panel with black boxes around their point. Only four energy-limited simulations strip the planet, with three of these occurring at late times ($>8$\,Gyr). These simulations, like those we presented in \citet{King2022,King2024}, highlight the possibility of late time evolution for some planets, even if it turns out not to be the case for \DSTA\,b. This is likely a result of the shallower EUV decline in the Mors code \citep{Johnstone2021} compared to prescriptions adopted in many earlier photoevaporation studies \citep[see also][]{EUVevolution}.

\subsection{Current Mass Loss}

\begin{table}[]
\centering
\caption{Calculated current mass loss rates for \DSTA\,b.}
\label{tab:curMassLoss}
\begin{tabular}{cccc}
\hline
Mass         & Energy-       & ATES          & ATES          \\
             & Limited       & Hydro.        & Approx        \\
(M$_\oplus$) & (g\,s$^{-1}$) & (g\,s$^{-1}$) & (g\,s$^{-1}$) \\ \hline
6            & $8.2 \times 10^{11}$        & n/a$^\dagger$ & $1.2 \times 10^{11}$        \\
10           & $4.7 \times 10^{11}$        & $1.7 \times 10^{11}$        & $8.5 \times 10^{11}$        \\
15           & $3.1 \times 10^{11}$        & $1.0 \times 10^{11}$        & $6.3 \times 10^{11}$        \\
20           & $2.3 \times 10^{11}$        & $7.6 \times 10^{11}$        & $5.2 \times 10^{11}$        \\ \hline
\end{tabular}
\tablecomments{$^\dagger$ The ATES code returned NaN for the 6\,M$_\oplus$ case.}
\end{table}

We calculated current mass loss estimates using three mass loss rate calculation methods: the two used for the lifetime simulations, as well as running the full hydrodynamic ATES code. We did these calculations with four different assumptions for the total mass of the planet: 6, 10, 15, and 20\,M$_\oplus$. Our results are shown in Table \ref{tab:curMassLoss}. When we set the planet mass to 6\,M$_\oplus$, the ATES code failed to converge on a solution and returned NaN. We assume this is due to the small mass for a planet of this size causing the code to struggle to reach a steady-state. 

All of the values suggest substantial mass loss is ongoing at the current time, as expected from the high XUV emission and young age of the system, and could translate into direct detection of evporation for this planet being possible. However, other considerations such as ionization level of the escaping material and how the UV irradiation balance affects the population of the metastable helium state must be factored in for determining detectability at Ly-$\alpha$ and the 10830\,\AA\ triplet, respectively. Simulations to perform such assessments are beyond the scope of this work. High levels of stellar variation on short timescales, as we observe for \DSTA\ in X-rays in Fig.~\ref{fig:xLC}, could also act to make detecting escaping material more difficult.

The ATES approximation mass loss rates are slightly below those for the hydrodynamic code itself, but always within a factor of two. This suggests that somewhat more stripping than shown in our simulations in Section \ref{ssec:Sims} might be possible, at least at early ages. The small factor of the differences however mean that the overall conclusions are unlikely to change much if it were possible to run the hydrodynamic code at each step of the simulations. The energy-limited results suggest that our choice of $\eta\beta^2 = 0.15$ for those simulations may be too low for this planet at the current state. Nevertheless, the canonical values we use for these parameters provide a means for easy comparison with past studies which made the same or similar assumptions \citep[e.g.][]{Ehrenreich2011,Salz2015,Erkaev2016,Owen2017,King2018,Kubyshkina2018}.

We note that \citet{Foster2022} estimated the mass loss rate of \DSTA\,b to be $6.05 \times 10^{11}$\,g\,s$^{-1}$ based on the \textit{eROSITA} measurement, also using energy-limited escape. This value agrees well with our ranges in Table~\ref{tab:curMassLoss}, despite their set of assumptions differing slightly from ours. They assumed a 50-50 contribution of the two stars to the unresolved X-ray detection, $\beta = 1.1$, ignored Roche lobe effects, and used an estimated mass of 26.7\,M$_\oplus$. This last value was based on the mass-radius relation of \citet{Chen2017}, and exceeds the 1-$\sigma$ upper limit by almost a factor of two.

\section{Conclusions}
\label{sec:conclusions}

Four observations of the DS\,Tuc system by \textit{Chandra} in 2022 have revealed X-ray variation of the two stars on multiple timescales.

Star A flared strongly twice, and both stars show further evidence of emission enhancement at various other points in the observations. From this, we estimated an X-ray flare rate of 2.7 per day, though the low numbers involved mean that it is uncertain how representative of the true flare rate this number is. We note though that it is in excellent agreement with the rate across previous \textit{XMM-Newton} observations analyzed by \citet{Benatti2021} and \citet{Pillitteri2022}. The rate also agrees with optical band predictions by \citet{Colombo2022}, though all four flares unambiguously identified so far are far in excess of their predicted flare energy. The small archival \textit{Swift} dataset also contains a possible flare, but the spatial resolution is too low to determine which star was responsible for the observed count rate increase.

The quiescent levels of both stars show significant variation between the four epochs. \DSTB's variation shows no clear pattern, and the first observation is much lower than the other three. \DSTA\ shows a clear, monotonic decline from each observation to the next, that could be due to a Solar-like activity cycle. No such cycle has been previously confirmed for a star this young, and this exciting hint motivates more observations in the coming years to explore this possibility further. 

We compared the spread of our measured $L_{\rm X}/L_{\rm bol}$ values across the observations and different activity states of the stars to the spread of $L_{\rm X}/L_{\rm bol}$ among stars of similar Rossby numbers. We found that for both stars the spread of points is statistically consistent with the field sample, suggesting that temporal variation could be a major contributor to the observed scatter in activity relationships.

In simulating the possible future of the planet \DSTA\,b, we found a wide range of possible scenarios from the planet ending up Neptune-sized all the way down to a super-Earth whose primordial H/He envelope is completely stripped. The latter outcome was more likely when we used the ATES method for mass loss estimation, with 63\% of tracks reproducing the current radius eventually resulting in a stripped planet. Interestingly, in the cases where the planet is fully stripped, half of these retain some portion of their envelope until after an age of 1\,Gyr, with a significant minority not completely stripped until well past 5\,Gyr. Better constraints on the mass of \DSTA\,b will enhance our predictions for its future by narrowing the number of possible scenarios.

We finally note that in advocating for further observation of DS\,Tuc in the future, \textit{Chandra}'s ACIS-S is the ideal current generation instrument to use. Despite the soft energy issues with the detector, the youth of the star means its X-ray emission peaks at harder energies around 1\,keV, significantly reducing the proportion of emission that is too soft for ACIS-S to detect. Meanwhile, there is a huge gain in precision from not having to estimate the contamination of the stars to each other, as their PSFs are fully separated.

\begin{acknowledgments}
We thank the anonymous referee for their quick reviews which helped improve this work. The scientific results reported in this article are based on observations made by the Chandra X-ray Observatory. This research has made use of software provided by the Chandra X-ray Center (CXC) in the application packages CIAO, ChIPS, and Sherpa. GK and LC acknowledge support for this work provided by the National Aeronautics and Space Administration through Chandra Award Number GO2-23001X issued by the Chandra X-ray Center, which is operated by the Smithsonian Astrophysical Observatory for and on behalf of the National Aeronautics Space Administration under contract NAS8-03060. PJW acknowledges support from the UK Science and Technology Facilities Council (STFC) through consolidated grants ST/T000406/1 and ST/X001121/1.
\end{acknowledgments}

%

\vspace{5mm}
\facilities{\textit{Chandra}, \textit{Swift}}


\software{astropy \citep{Astropy},
          CIAO \citep{ciao},
          Xspec \citep{xspec}
          }






\bibliography{DSTuc}{}
\bibliographystyle{aasjournal}



\end{document}